\documentclass[conference]{IEEEtran}
\usepackage{tabularx}
\usepackage{tabu}
\usepackage{blindtext}
\usepackage{multirow}
\usepackage{algpseudocode}
\usepackage{algorithm}
\usepackage{graphicx}
\usepackage{url}
\usepackage[T1]{fontenc}
\usepackage{tikz}
\usepackage{listings}
\usepackage{bm,times}
\usepackage{amsthm}
\usepackage{textcomp}
\usepackage{mathtools}
\usepackage{amsmath,amssymb}

\usepackage[numbers]{natbib}
\usepackage{caption}
\usepackage{subfigure}
\usepackage{bm}

\DeclarePairedDelimiter\abs{\lvert}{\rvert}%
\DeclarePairedDelimiterX\Set[2]{\lbrace}{\rbrace}%
{ #1 \,\delimsize|\,\mathopen{} #2 }
\newsavebox{\ieeealgbox}

\ifCLASSINFOpdf
\else
\fi
\begin{document}
%
\title{OptEx: A Deadline-Aware Cost Optimization Model for Spark}

\author{\IEEEauthorblockN{Subhajit Sidhanta\IEEEauthorrefmark{1}, Wojciech Golab\IEEEauthorrefmark{2}, and Supratik Mukhopadhyay\IEEEauthorrefmark{1}}
\IEEEauthorblockA{\IEEEauthorrefmark{1}Louisiana State University, Baton Rouge, Louisiana, USA, Email: \{ssidha1, supratik\}@csc.lsu.edu}
\IEEEauthorblockA{\IEEEauthorrefmark{2}University of Waterloo, Waterloo, Ontario, Canada, Email:  wgolab@uwaterloo.ca}}


%


\maketitle
\thispagestyle{plain}
\pagestyle{plain}

\begin{abstract}
 We present OptEx, a closed-form model of job execution on Apache
 Spark, a popular parallel processing engine. 
 To the best of our knowledge,
 OptEx is the first work that analytically models job completion time on Spark. 
  The model can be used to estimate the completion time of a given Spark job on a cloud, with respect to the size of the input dataset, the number of iterations, the number of nodes comprising the underlying cluster.
  Experimental results demonstrate that OptEx yields
a mean relative error of 6\%
 in estimating the job completion time.  
  Furthermore, the model can be applied for estimating
  the cost optimal cluster composition for running a given 
  Spark job on a cloud under a completion deadline
  specified in the  \emph{SLO} (i.e., Service Level Objective). 
   We show experimentally that OptEx is able to correctly
  estimate the cost optimal cluster composition for running a given Spark job under an SLO deadline with
  an accuracy of 98\%.  
\end{abstract}


%
\IEEEpeerreviewmaketitle

\section{Introduction}
Optimizing the cost of usage of cloud resources for running data-intensive jobs on large-scale parallel
processing engines is an important, yet relatively less explored problem.
   Cloud service providers, like Amazon, Rackspace, Microsoft, etc., allow users to outsource the hosting of applications and services to a cloud using clusters of \emph{virtual machine instances}. The cloud service providers charge a \emph{service usage cost} to the user on the basis of the hourly usage \cite{inc_amazon_2008} of the virtual machine instances.
   The cloud service providers present the users with a variety of
 virtual machine instance types to choose from, such as  micro, small, large, etc.,
 for Amazon Ec2  \cite{inc_amazon_2008}. Each virtual machine instance type has a different specification, in terms of CPU, I/O, etc.,  and different hourly usage cost. 
  The \emph{cost optimal cluster composition} 
 specifies a number of virtual machine instances (of different virtual
 machine instance types), that enable execution of the given job under the SLO (i.e., Service Level Objective) deadline, while  minimizing the service usage cost. However, the current state-of-the-art \cite{daniel2011prediction, Imai:2013:ARP:2588611.2588688, singer2010towards} cluster provisioning
 solutions do not ensure that a given SLO deadline for job execution is
 satisfied, while at the same time above service usage cost is minimized.
   \par We present OptEx\footnote{The project is partially supported by Army Research Office (ARO) under Grant W911\-NF1010495.  Any opinions, findings, and conclusions or recommendations expressed in this material are those of the authors and do not necessarily reflect
the views of the ARO or the United States Government.},
a closed-form job execution model for Apache Spark  \cite{Zaharia:2012:RDD:2228298.2228301}, a popular
 parallel processing engine. OptEx can be used to
determine  the  cost optimal cluster composition, comprising virtual machine instances provided by cloud service providers, like Amazon, RackSpace, Microsoft, etc., 
  for executing a given Spark job under an SLO deadline.
  As far as we know, OptEx
is the first work that analytically models job execution on Spark.
  OptEx analytically models the \emph{job completion time} of Spark jobs on a cluster of virtual machine
 instances. It decomposes the
 execution of  a target Spark job into smaller phases
, and models the completion time of each phase in terms of: 1) 
 the cluster size, the number of iterations,
   the input dataset size, and 2) certain model parameters estimated using \emph{job profiles}. OptEx categorizes Spark applications into
  application categories, and generates separate job profiles for each application category by executing
  specific \emph{representative jobs}. 
  The model parameters for the target job are estimated from the components
  of the job profile, corresponding to the application category of the target job.
  Experimental results demonstrate that OptEx yields
a mean relative error of 6\% in
 estimating the job completion time. 
\par  
 Using the model of job completion time (OptEx), we derive the objective
 function for minimizing the service usage cost for running a given Spark job under an SLO deadline.
   The cost optimal cluster composition for running the target Spark job under the SLO deadline is obtained using constrained
    optimization on the above objective function.
 Experimental results demonstrate that OptEx is able to correctly
  estimate the cost optimal cluster composition for running a given Spark job under an SLO deadline with
  an accuracy of 98\%.
  We also demonstrate experimentally that OptEx can be used to design
an optimal schedule for running a given job on a given cluster composition under an SLO deadline. 
 \par 
 Consider the use case where a web development company needs to run a Spark PageRank 
 application to determine the most important web pages they developed over the years, using the infrastructure (cluster)
 provided by a popular cloud provider, like Amazon. Using state-of-the-art \cite{daniel2011prediction, Imai:2013:ARP:2588611.2588688, singer2010towards}
 prior experience-based provisioning techniques
, the company may provision a  cluster of 30 m2.xlarge Amazon Ec2 instances
 to run the Spark job, under an SLO deadline of 70 hours. In this case,  they may end up actually
finishing the job in 40 hours, incurring a service usage cost of \$168.45 (at the hourly rate of 0.1403 using the pricing
scheme from \cite{inc_amazon_2008}).
 However, with OptEx, 
 the job would have completed in 60 hours using only 10 m2.xlarge Amazon Ec2 nodes, incurring just \$84.18,
 while satisfying the deadline. Thus, OptEx helps minimizing the service
 usage cost without violating the SLO deadline.
  \par The technical contributions of this paper are summarized as follows.
  \begin{itemize}
  \item  We present OptEx, an analytical model for Spark \cite{Zaharia:2012:RDD:2228298.2228301} job execution. 
 \item We provide a technique for estimating the cost optimal cluster composition for running a given Spark job under an SLO deadline, using the above model.
  \end{itemize}

 \subsection{Motivation}\label{sec:costcase}
 Scaling out (i.e., adding nodes
to the cluster) \cite{Herodotou:2011:NOS:2038916.2038934} is the common way of increasing the performance of  parallel processing on the cloud. 
 Cost is an important factor in scaling out, with the cost of cluster usage increasing linearly with the number of virtual machine instances in a cluster, evident from analysis of the Amazon Ec2 pricing policy \cite{inc_amazon_2008}. Hence for minimizing the service usage cost, determining the optimal cluster size for executing a given job is of utmost importance. 
  In the current state-of-the-art \cite{daniel2011prediction, Imai:2013:ARP:2588611.2588688, singer2010towards}, a cloud service consumer can choose the required cluster configuration in one of the following manners: 1) arbitrarily, or 2) make an informed decision using previous experience of running similar jobs on the cloud. 
 However none of the above strategies ensure that the SLO deadline is satisfied, and at the same time the service usage cost is minimized. Elastisizer \cite{Herodotou:2011:NOS:2038916.2038934} is the only successful work in this direction, but it addresses Hadoop MapReduce and does not address Spark. 

 \subsection{The OptEx Approach}\label{sec:modelapproach}
 OptEx decomposes a Spark job execution into different phases, namely
 the initialization phase, the preparation phase, the variable sharing phase, and the computation phase.
 Following an analytical modelling approach, OptEx expresses the execution time of each phase in terms of the cluster size, number of iterations, the input dataset size, and certain model parameters. Similar to the ARIA framework \cite{Verma:2011:AAR:1998582.1998637}, which applies profiling for scheduling Hadoop MapReduce jobs, OptEx estimates the model parameters with the components of the specific job
 profile corresponding to the application category  of the target job. 
 ARIA uses Hadoop-specific parameters for profiling \cite{Verma:2011:AAR:1998582.1998637}, and hence is unsuitable for application to Spark.
  OptEx considers job completion deadline as an SLO parameter \cite{Chi:2011:IIC:2002938.2002942, Zhang:2011:TCS:2095686.2095687}, 
 that acts as the constraint for minimizing the service usage cost. 
  An objective function for the service usage cost is obtained based on the job execution model. Constrained optimization techniques \cite{leader2004numerical} are applied on the objective
   function to estimate the cost optimal cluster composition for finishing a Spark job within a given SLO deadline.

\section{Spark Job Execution Phases}\label{sec:exec}
\begin{figure}[!ht]
 \centering
 \captionsetup{justification=centering}
 \includegraphics[width=3in,height=1in]
 {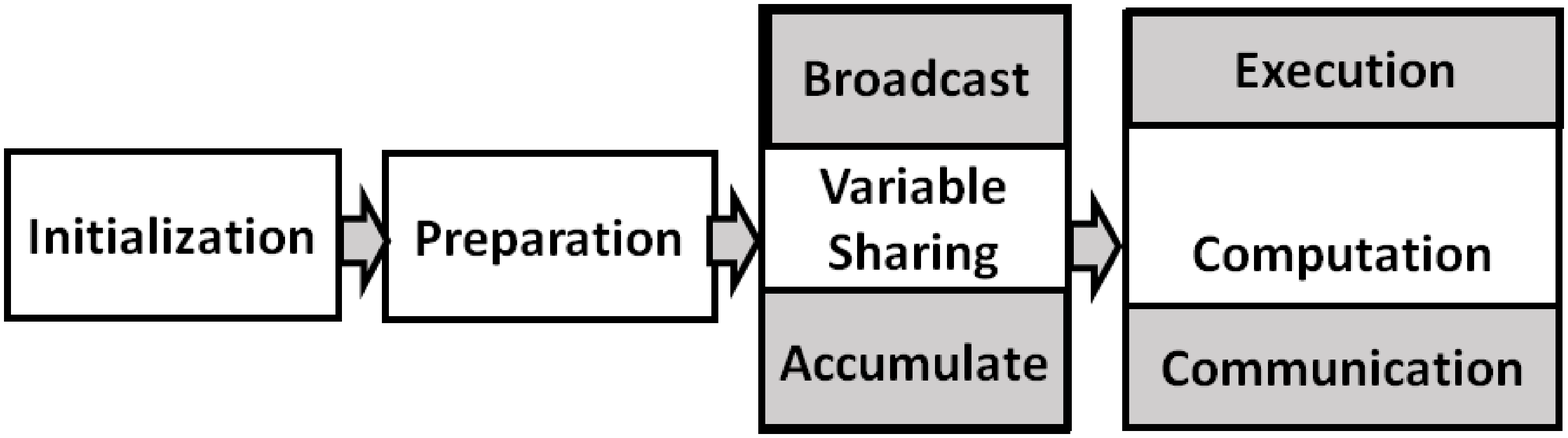}
 \caption{Phases in a Spark Job Execution Flow}
 \label{fig:phases}
 \end{figure}
  We decompose a typical Spark job execution flow into logically distinct phases illustrated in Figure \ref{fig:phases}. Each of these phases behave differently
with respect to variations in the number of iterations, the cluster size, and the dataset size. The first phase in a Spark job is the \emph{initialization}
phase, which performs activities like class loading, symbol table creation, object initialization, function
loading, and logger initialization. 
 The second phase is the
\emph{preparation phase}, which is responsible for job scheduling, resource allocation,
and context creation. The initialization and preparation phases
are relatively invariant to changes in input variables \cite{Zaharia:2012:RDD:2228298.2228301}. 
 The next phase is the \emph{variable sharing phase} that deals with broadcasting or accumulating blocks of data from the Spark master to the workers.
\par Spark uses a novel in-memory data structure called the
 \emph{RDD} (i.e., resilient distributed dataset) for fast and fault tolerant computation \cite{Zaharia:2012:RDD:2228298.2228301}. Internally,
 each Spark job is processed as a permutation of several \emph{unit RDD tasks} (operations like flatmap, reduce, etc),
  that are executed in parallel
  on the worker nodes. Spark provides a wide range of built-in unit RDD tasks, packaged within several library modules \cite{apache:library1234},
  like MLlib, Spark SQL modules, etc. During the last phase, i.e., the \emph{computation phase} (Figure \ref{fig:phases}), the given application makes calls
to methods from the above library modules, which in turn triggers the respective unit RDD tasks on the workers. 
 The computation phase comprises: 1) the \emph{communication phase} that communicates the intermediate variables among the workers,
and 2) the \emph{execution phase} that involves the actual execution of the unit RDD tasks on the workers. The lengths of the variable sharing phase and the computation phase monotonically increase
with the input variables, i.e., the number of iterations, the cluster size, and the dataset size \cite{Zaharia:2012:RDD:2228298.2228301}. In particular, the variable sharing phase and the computation phase are repeated under iterations, and the lengths of the above phases increase with respect to number of iterations.  

\section{Application of Profiling For Estimating the Model Parameters} \label{sec:profile}
 The most common technique for estimating the performance of a given job \cite{Verma:2011:AAR:1998582.1998637,
brebner2011} is using a standard profiling tool, that measures real-time performance statistics, to generate job
 profile for the target job using a representative
job. OptEx categorizes Spark applications,
 and uses profiling to
 generate separate job profiles for 
  each application category with representative jobs for each category.
   Components of the job
 profile are used as estimates for the model parameters of the target job. 

 \subsection{Application Categorization}\label{sec:cat} 
   As discussed earlier, OptEx categorizes Spark applications into application categories, chooses a representative
   job for each category, and creates the job profile for that category using the respective representative job.
    Application categorization is a difficult open problem, dependent on the application domain, and beyond
    the scope of this paper. OptEx allows the developers to choose their own
    categorization scheme.  
   \par In this paper, we categorize Spark applications
   according to the category of library modules that an application uses. Each Spark application uses specific libraries 
   depending on the business logic \cite{Zaharia:2012:RDD:2228298.2228301}. The Apache Spark distribution \cite{apache:library1234}
   currently organizes the library modules into following four categories: 1) Spark SQL, which supports Apache Hive 
    or JDBC 
     queries, 2) Spark Streaming, which comprises streaming applications, 3) MLlib, which supports machine
     learning, and 4) GraphX, which facilitates working with graphs and collections. Thus, OptEx uses four application
      categories, and a specific job profile for each category is obtained using a representative job for each
      category.
 \subsection{Choice of Representative Jobs For Each Category} \label{sec:rep}
     The execution phase (Figure \ref{fig:phases}) of a Spark job comprises a permutation of low-level unit RDD tasks. 
 We call an application $a$ to be a representative job  for a given job $j$, if:
    1) the job $a$ contains all the
 unit RDD tasks comprising the job $j$,
 and 2) if job $j$ is iterative, $a$ is also iterative, and vice versa. 
      According to the given categorization scheme (Section \ref{sec:cat}), OptEx categorizes applications
     on the basis of the Spark library modules they use. The Apache Spark distribution web page \cite{apache:library1234} describes an
     example application for each group of library modules. For a given application category, the respective example application is chosen as the representative job, under the given categorization scheme. 
      By the design of the Spark libraries, these chosen jobs trivially satisfy the above two conditions for being a representative job.
     \par The custom Spark application, mentioned as an example
     in the web page of Spark Streaming library \cite{apache:library1234}, is used as the representative job for the applications using the Spark Streaming.
     It runs on the Twitter dataset \cite{Zafarani+Liu:2009}
, and lists the current tweets on a sliding window. 
  Similarly, the representative application for the MLlib group of applications is the movie rating application MovieLensALS \cite{apache:library1234}. 
  The input workload for the MovieLensALS applications is the
MovieLens dataset made available by Netflix at grouplens.org \cite{MOVIELENS-DATA}. 
 PageRank 
  is the representative application for the GraphX group of applications \cite{apache:library1234}.
 In the absence of an example application for the Spark SQL category, the widely used Big Data Benchmark \cite{amplab:benchmark7890} developed by AMPLab is used as the representative job for this category. It has been widely accepted as a benchmark for Spark SQL jobs \cite{apache:library1234}. 
\newtheorem{definition}{Definition}
\newcommand{\argmax}{\arg\!\max}
\newcommand{\card}[1]{\ensuremath{\left\|#1\right\|}}
\def\tuple#1{\langle #1\rangle}
\newcommand{\bigslant}[2]{{\raisebox{.2em}{$#1$}\left/\raisebox{-.2em}{$#2$}\right.}}

\subsection{Estimation of Model Parameters from the Job Profile}\label{sec:appprof}
\begin{table}[h]
\scriptsize
\caption{Glossary of symbols and terms} 
\centering 
\scalebox{0.9}{
\begin{tabular}{|l|l|l|l|l|}
\hline
$\mathit{T_{\mathit{vs}}}$                     & \begin{tabular}[x]{@{}l@{}}Estimated completion \\ time for the variable \\ sharing phase\end{tabular}                                                        &                    & $\mathit{T_{\mathit{Est}}}$         & \begin{tabular}[x]{@{}l@{}}Estimated job \\ completion time\end{tabular}                                              \\ \cline{1-2} \cline{4-5}
$n$                        & \begin{tabular}[x]{@{}l@{}}The cluster size\end{tabular}                                                       &                    & $\mathit{T_{\mathit{init}}}$             & \begin{tabular}[x]{@{}l@{}}Estimated completion \\ time for the \\ initialization phase\end{tabular}                                           \\ \cline{1-2} \cline{4-5}
$\mathit{M_{a}^k}$ & \begin{tabular}[x]{@{}l@{}}Execution time of \\ the $\mathit{k^{th}}$ RDD operation \\ for job $a$\end{tabular}                         &                    & $\mathit{T_{\mathit{prep}}}$     & \begin{tabular}[x]{@{}l@{}}Estimated completion \\ time for the \\ preparation phase\end{tabular}                       \\ \cline{1-2} \cline{4-5}
$\mathit{T_{\mathit{Rec}}}$                     & \begin{tabular}[x]{@{}l@{}} Recorded execution \\ time\end{tabular}                                                    &                    & $\mathit{iter}$          & \begin{tabular}[x]{@{}l@{}}Number of iterations \end{tabular}                                                     \\ \cline{1-2} \cline{4-5}
t                        & \begin{tabular}[x]{@{}l@{}}number of possible \\ instance types\end{tabular}                                                        &                    & $s$          & \begin{tabular}[x]{@{}l@{}}size of input dataset\end{tabular}                                                     \\ \cline{1-2} \cline{4-5}
$\mathit{T_{\mathit{comp}}}$                     & \begin{tabular}[x]{@{}l@{}}Estimated completion \\ time for the \\ computation phase\end{tabular}                                                 &                    & $\mathit{T_{\mathit{commn}}}$             & \begin{tabular}[x]{@{}l@{}}Estimated completion \\ time for the \\ communication phase \\ in $\mathit{T_{\mathit{comp}}}$\end{tabular}                                      \\ \hline
$T_{vs}^{\mathit{baseline}}$                     & \begin{tabular}[x]{@{}l@{}} baseline value of \\ $\mathit{T_{\mathit{vs}}}$ \end{tabular}                                                 &                    & $\mathit{coeff}$             & \begin{tabular}[x]{@{}l@{}}coefficient of $\mathit{T_{\mathit{vs}}}$ \\ in $\mathit{T_{\mathit{Est}}}$\end{tabular}                                       \\ \hline
$\mathit{T_{\mathit{commn}}^{\mathit{baseline}}}$                     & \begin{tabular}[x]{@{}l@{}}baseline value of \\ $\mathit{T_{\mathit{commn}}}$\end{tabular}                                                 &                    &   \begin{tabular}[x]{@{}l@{}}$\mathit{cf_{\mathit{commn}}}$\end{tabular}          & \begin{tabular}[x]{@{}l@{}}coefficient of \\ $\mathit{T_{\mathit{commn}}}$ in $\mathit{T_{Est}}$\end{tabular}                                     \\ \hline
\end{tabular}}
\label{table:glossary}
\end{table}

\begin{table}[!h]
  \caption{An Example Job Profile: Profile for MLlib jobs on m1.large instances} 
\centering 
\resizebox{\columnwidth}{!}{%
\begin{tabular}{|c|l|l|l|l|l|l|l|l|}
\hline
\multicolumn{1}{l}{App} & $T_{\mathit{init}}$(sec)     & $T_{\mathit{prep}}$(sec)     & $T_{vs}^{baseline}$(sec) & $\mathit{coeff}$               & $T_{\mathit{commn}}^{\mathit{baseline}}$(sec) & $\mathit{cf_{\mathit{commn}}}$         & \multicolumn{2}{c}{$T_{\mathit{exec}}$} \\
\hline
\multirow{7}{*}{ALS}    & \multirow{7}{*}{20} & \multirow{7}{*}{13} & \multirow{7}{*}{15}       & \multirow{7}{*}{0.004} & \multirow{7}{*}{11}         & \multirow{7}{*}{0.07} & RDD task  & $M_{a}^{k}$(ms)  \\
\hline
                        &                     &                     &                           &                       &                             &                       & mean      & 100              \\
                        &                     &                     &                           &                       &                             &                       & map       & 98               \\
                        &                     &                     &                           &                       &                             &                       & flatmap   & 72               \\
                        &                     &                     &                           &                       &                             &                       & first     & 5                \\
                        &                     &                     &                           &                       &                             &                       & count     & 124              \\
                        &                     &                     &                           &                       &                             &                       & distinct  & 300
                        \\ \hline
\end{tabular}
}
\label{table:profile}
\end{table}

   A Spark 
   job is typically written in a high-level language (like Scala, Python, etc.) internally executed in different phases (Figure
   \ref{fig:phases}). The length of the initialization phase $\mathit{T_{\mathit{Init}}}$ and the length of the
  preparation phase ${T_{\mathit{prep}}}$ remain constant to variations in the input variables \cite{Zaharia:2012:RDD:2228298.2228301}. The length of
  the execution phase (${T_{\mathit{exec}}}$) and the length of the variable
    sharing phase  (${T_{\mathit{vs}}}$) increase monotonically with respect to the input variables \cite{Zaharia:2012:RDD:2228298.2228301}. Thus, the length
    of these phases in the execution of a representative job (contained in the job profile) can act as the point of reference, i.e., \emph{baseline}, for measuring
    the length of the corresponding phases in the target job.  
    In this section, we elaborate how these baseline values in the job profile
   can be used for estimating the parameters of the model for each job phase. 
   \par 
    During profiling,
 the representative application
 $a$ is run on a single node, and the length of the initialization phase, the preparation phase, the variable sharing phase, and the communication phase (Figure \ref{fig:phases}) is recorded in the job profile. The length  of the
 above phases in the job profile act as baseline values for estimating the lengths of the corresponding phases in a given target job.  
The length of the initialization phase $\mathit{T_{\mathit{Init}}}$ (Table \ref{table:profile}) and the length of the
  preparation phase ${T_{\mathit{prep}}}$ for a given job are directly estimated from the lengths of the corresponding phases in the job
  profile (since, as discussed in Section \ref{sec:exec}, these phases remain constant with respect to the variations in the input variables).
  As elaborated in Section \ref{sec:exec}, the length of the variable sharing phase $\mathit{T_{\mathit{vs}}}$ increases monotonically \cite{Zaharia:2012:RDD:2228298.2228301}
  with respect to the cluster size and the number of iterations. Hence, $\mathit{T_{\mathit{vs}}}$ is expressed as a function of:
 \begin{itemize}
 \item The input   variable $n$ represents
 the number of nodes. 
 \item  The input variable $\mathit{iter}$ represents the number of iterations. 
 \item  The baseline value $\mathit{T_{\mathit{vs}}^{\mathit{baseline}}}$, contained in the job profile, representing the length of the
 variable sharing phase of the representative application. It is the
 baseline for estimating the length of the variable sharing phase $\mathit{T_{\mathit{vs}}}$ of a given target job. 
\end{itemize}
 The increase in the length of the variable sharing phase $\mathit{T_{\mathit{vs}}}$, estimated relative to the
 baseline
 $\mathit{T_{\mathit{vs}}^{\mathit{baseline}}}$,  in terms of the given values of the input variables $n$ and
 $\mathit{iter}$ (we compute the total duration of the variable sharing phase across all iterations), i.e., $\mathit{T_{vs}}$ is expressed as:  
   \begin{equation}\label{eqn:brd}
   \mathit{T_{\mathit{vs}}} = \mathit{coeff} \times \mathit{iter} \times n \times \mathit{T_{\mathit{vs}}^{\mathit{baseline}}},
   \end{equation}
   where $\mathit{iter}$ is the number of iterations,
   $n$ is the number of nodes, $\mathit{T_{\mathit{vs}}^{\mathit{baseline}}}$ is the baseline value, and $\mathit{coeff}$ is a coefficient term.
 The coefficient term $\mathit{coeff}$  is empirically estimated during job profiling 
 using curve fitting on the results of repetitive experiments with the representative job.
  The length of the computation phase $\mathit{T_{\mathit{comp}}}$ (Table \ref{table:glossary})
  is made up of two logical components:
 the length of the communication phase $\mathit{T_{\mathit{commn}}}$, and the length of the execution phase
 $\mathit{T_{\mathit{exec}}}$ (Figure \ref{fig:phases}). 
   \par The communication phase is responsible for fetching the
 values of the intermediate variables computed by tasks in the
 earlier stages of the given job. While profiling, the length of the communication phase of a representative
 application $a$ on a single node is recorded in the job profile as $\mathit{T_{\mathit{commn}}^{\mathit{baseline}}}$. It serves as the baseline measure against
  which the length the communication phase $\mathit{T_{commn}}$ is estimated. 
  The size $s$ of the input dataset is given in bytes (for example, the size of the input for the wordcount application is
 given as the size of the input files). Since the length of the communication phase $\mathit{T_{commn}}$ increases with respect to the input variable $s$ \cite{Zaharia:2012:RDD:2228298.2228301},
  $\mathit{T_{commn}}$ is expressed as a product
of the input dataset size $s$, a coefficient $\mathit{\mathit{cf}_{\mathit{commn}}}$, and the baseline value
$\mathit{T_{\mathit{commn}}^{baseline}}$, where $\mathit{\mathit{cf}_{\mathit{commn}}}$  and $\mathit{T_{\mathit{commn}}^{\mathit{baseline}}}$ are
obtained from the job profile. Again, the coefficient $\mathit{\mathit{cf}_{commn}}$ is empirically estimated in the
profiling stage applying curve fitting on the outputs of experiments with the representative job.
 Thus, 
 \begin{equation}\label{eqn:commn2}
 \mathit{T_{\mathit{commn}}} = \mathit{\mathit{cf}_{\mathit{commn}}} \times  \mathit{T_{\mathit{commn}}^{\mathit{baseline}}} \times \mathit{s}
 \end{equation}
  \par As discussed in Section \ref{sec:rep}, the execution phase of a Spark job comprises a permutation
 of unit RDD tasks. 
    The OptEx job profile records the average running time
$\mathit{M_{a}^{k}}$ (Table \ref{table:glossary}) of each unit RDD task component $k$ comprising the representative Spark application $a$. If there are multiple occurrences of an RDD task $i$ in $a$, we consider the
average running time for all occurrences of the task $i$. By design the representative job for an application category
contains all the unit RDD tasks comprising any given job in that category. 
 Hence, the
length of the execution phase of the given Spark job is estimated as a function of the average running time $\mathit{M_{a}^{k}}$
 of each unit RDD task $k$ in the job profile.

 \section{Derivation of the Spark Job Execution Model}\label{sec:model}
 \subsection{Input Variables} \label{sec:var}
 OptEx accepts the following input variables:
   the size $s$ of the input dataset in bytes, 
   the number of nodes $n$ in the cluster, and
   the number of iterations $\mathit{iter}$ in the given job \cite{Zaharia:2012:RDD:2228298.2228301}.
    The
number of iterations for an iterative Spark application is typically passed as a runtime argument by the developer
\cite{Zaharia:2012:RDD:2228298.2228301}. Moreover, Spark applications typically have only  few lines of code. 
 Hence if we need to determine $iter$ from the code, we do not require sophisticated techniques involving static analysis \cite{Zuleger:2011:BAI:2041552.2041574}
. The other input variables, i.e., number of nodes $n$
and input dataset size $s$, are also directly
 passed to the model as runtime arguments.
 \par  While modelling the estimated total completion time for the target job, the user provides an
   estimated upper bound for the number of iterations $\mathit{iter}$ for the target job, as an input to the
   model. During the actual running time of the target job, the user provides the number of iterations
   $\mathit{\mathit{iter}_{exec}}$ as a
   runtime argument to the job \cite{Zaharia:2012:RDD:2228298.2228301}.
   The number of iterations $\mathit{\mathit{iter}_{exec}}$ provided in the running time may differ from
    the number of iterations $\mathit{iter}$ provided in the modelling phase.  The difference between
   $\mathit{\mathit{iter}_{exec}}$ and $\mathit{iter}$ may cause: 1) unpredicted
  wastage of cluster resources, and 2) the failure to satisfy the SLO.
  \newcommand{\BigTheta}[1]{\ensuremath{\operatorname{\Theta}\bigl(#1\bigr)}}
   In that case, the estimations need to be  redone, with a new input value 
    for the number of iterations. For multiple runs of the target job with different values of the runtime argument
  $\mathit{\mathit{iter}_{exec}}$ supplied by the user in each run, the maximum of the $\mathit{\mathit{iter}_{exec}}$
  values, i.e., $\mathit{\mathit{iter}_{\mathit{exec}}^{\mathit{max}}}$, is supplied as the new input for the estimation.
   The estimation using the new value $\mathit{\mathit{iter}_{\mathit{exec}}^{\mathit{max}}}$ amounts to computing the value of $\mathit{T_{\mathit{Est}}}$ from
     the Equation \ref{eqn:est2} with a time complexity of \BigTheta{1} (since the degree of
     $\mathit{T_{\mathit{Est}}}$ is 1 \cite{leader2004numerical}), thus incurring negligible overhead.

 \subsection{Formulation of the Model}\label{sec:formulation}
OptEx decomposes the job completion time into four phases (Figure \ref{fig:phases}), and
models the total job completion time $\mathit{T_{\mathit{Est}}}$ as the sum of the lengths of the component phases. 
Thus, \begin{equation}\label{eqn:est1}
 \mathit{T_{\mathit{Est}}}  =   \mathit{T_{\mathit{Init}}}  +    \mathit{T_{\mathit{prep}}} +  \mathit{T_{\mathit{vs}}}  +  \mathit{T_{\mathit{comp}}},
 \end{equation} where $\mathit{T_{\mathit{Init}}}$ is the length of the initialization phase, ${T_{\mathit{prep}}}$ is  the length of the preparation phase, $\mathit{T_{\mathit{vs}}}$ is the length of the variable sharing phase, and $\mathit{T_{\mathit{comp}}}$  is the length of the computation phase.
     As
   discussed in Section \ref{sec:rep}, 
 the execution phase of a given Spark job comprises a permutation
 of low-level unit RDD tasks. The number  of unit RDD tasks $\mathit{n_{unit}}$ increases monotonically with increasing
 the input dataset size $s$ and the number of iterations $\mathit{iter}$ \cite{Zaharia:2012:RDD:2228298.2228301}.
  Hence, the number  of unit RDD tasks $\mathit{n_{unit}}$
   can be expressed as a function comprising the following terms:
 \begin{itemize}
   \item  The size of the
   input dataset is denoted as $s$.
  \item The number of iterations  in the job is given as $\mathit{iter}$.
 \item The baseline term for the number of unit RDD tasks is given as $\mathit{n_{\mathit{unit}}^{\mathit{baseline}}}$. It is obtained from the job
 profile (Section \ref{sec:appprof}). Spark enables parallel execution by dividing the input dataset into partitions, and distributing the partitions/slices among the worker nodes  \cite{Zaharia:2012:RDD:2228298.2228301}.  $\mathit{n_{\mathit{unit}}^{\mathit{baseline}}}$ directly corresponds to the number of partitions that the input dataset is comprised of.
 The number of
 partitions can be: 1) computed from the size $s$ of the input dataset and the number of iterations $\mathit{iter}$ \cite{Zaharia:2012:RDD:2228298.2228301}, or
 2) programmatically provided as a parameter to the built-in transformation method used to create the RDDs
 from the input dataset \cite{Zaharia:2012:RDD:2228298.2228301}.
 \par For example, the Spark Wordcount program, working on input files from a HDFS backend,
 divides the input dataset into as many partitions as the number of HDFS blocks comprising the input files. Consider 
 a Wikipedia dump \cite{snapnets} consisting of 164 files, where the
 size of each file is less than the HDFS block size. Hence the number of partitions, and in turn the number
 of unit RDD tasks is 164. Thus, the baseline $\mathit{n_{\mathit{unit}}^{\mathit{baseline}}}$ is 164.

  \end{itemize}
 Thus, the increase of  $\mathit{n_{unit}}$, with respect to the above baseline $\mathit{n_{\mathit{unit}}^{\mathit{baseline}}}$, in terms
  of the parameters $s$ and $\mathit{iter}$, is expressed as
 \begin{equation}\label{eqn:s}
 \mathit{n_{unit}} = \mathit{n_{\mathit{unit}}^{baseline}} \times s \times \mathit{iter}.
 \end{equation}
 \par As discussed already in Section \ref{sec:appprof}, the length of the initialization phase $\mathit{T_{\mathit{Init}}}$ and the length of the
  preparation phase ${T_{\mathit{prep}}}$ are directly estimated from the corresponding components in the job
  profile (Section \ref{sec:appprof}). As discussed in Section \ref{sec:exec}, the length of the
  variable sharing phase $\mathit{T_{\mathit{vs}}}$ and the length of the computation phase $\mathit{T_{\mathit{commn}}}$
  vary with respect to the input variables. 
   Hence
  the length of the variable sharing phase $\mathit{T_{\mathit{vs}}}$ (Equation \ref{eqn:brd})
  and the length of the computation phase $\mathit{T_{\mathit{comp}}}$ are estimated as functions of the job profile components,
   and the input variables 
    (Section
  \ref{sec:appprof}). The expression for the length of the variable sharing phase $\mathit{T_{\mathit{vs}}}$, comprising the baseline value
 $\mathit{T_{\mathit{vs}}^{\mathit{baseline}}}$ and coefficient $\mathit{\mathit{coeff}}$ obtained from the job profile, is  given by
  Equation \ref{eqn:brd}.
  \par The length of the computation phase $\mathit{T_{\mathit{comp}}}$ in Equation
  \ref{eqn:est1} can be further decomposed into the following two logical components:
 \textbf{A)} \bm{$\mathit{T_{\mathit{\bm{commn}}}}$}: 
  The length of the communication phase $\mathit{T_{\mathit{commn}}}$ is obtained from the Equation \ref{eqn:commn2}. 
 \textbf{B)} \bm{$\mathit{T_{\mathit{\bm{exec}}}}$}: The length of the execution
 phase $\mathit{T_{\mathit{exec}}}$ in Equation \ref{eqn:est1} comprises the actual execution of $k$ RDD operations
 comprising the job on the worker nodes (Section \ref{sec:appprof}). $\mathit{T_{\mathit{exec}}}$ depends on various factors \cite{Zaharia:2012:RDD:2228298.2228301}: 1) the running times of the unit RDD tasks
 comprising the given job, 2) the number of iterations $\mathit{iter}$, 3) the number of stages in the job,
 4) parallelization of  the job across the worker nodes, and 5) sharing of the RDD variables across the cluster. Hence, execution phase length $\mathit{T_{\mathit{exec}}}$ is expressed as the sum  over
 the estimated computation times of all unit RDD tasks comprising j, along with coefficients accounting for the
 above factors.
 Thus, the length of the execution phase $\mathit{T_{\mathit{exec}}}$ of job $a$, without taking into account the parallelization factor $n$, is given as: 
  \begin{equation}\label{eqn:exec1}
  \mathit{T_{\mathit{exec}}} = \mathit{iter} \times  \mathit{\sum_{\substack{k=1}}^{\mathit{n_{unit}}}
                                        M_{a}^{k}},
  \end{equation}
   where  $\mathit{n_{unit}}$ is the number of unit RDD tasks given in Equation \ref{eqn:s}, $\mathit{M_a^k}$ is the average job execution
  time of a unit RDD task $k$ comprising the job $a$, and $\mathit{iter}$ is the number of iterations in the job.
  \par 
   Following prior work on modelling execution of parallel tasks \cite{Verma:2011:AAR:1998582.1998637}, the overall length of the computation phase $\mathit{T_{\mathit{comp}}}$ is divided by the factor $n$, taking into account
  the parallelization of the  across the $n$ worker nodes. Thus, the computation
  phase is rewritten as the sum of its two components, divided by $n$:
  \begin{equation}\label{eqn:comp2}
  \mathit{T_{\mathit{comp}}} = \frac{\mathit{T_{\mathit{commn}}} + \mathit{T_{\mathit{exec}}}}{n}.
  \end{equation}
Combining the Equations \ref{eqn:commn2}, \ref{eqn:exec1}, and \ref{eqn:comp2}, we get
\begin{align}\label{eqn:comp3}
\begin{split}
\mathit{T_{\mathit{comp}} = \mathit{iter} \times   \sum_{\substack{k=1}}^{\mathit{n_{unit}}} M_{a}^{k} / n}  + \frac{A \times s}{n},
\end{split}
\end{align} where $\mathit{n_{unit}}$ is the number of unit RDD tasks given in Equation \ref{eqn:s},
 and $A$ = $\mathit{\frac{\mathit{cf}_{\mathit{commn}} \times T_{\mathit{commn}}^{\mathit{baseline}}}{\mathit{s_{\mathit{baseline}}}}}$.
\begin{table}[!htpb]
  \caption{Stepwise Accuracy of Estimations}
\centering 
\scalebox{0.8}{
\begin{tabular}{|l|l|l|l|l|l|l|l|l|l|l|l|}
\hline
$iter$ & $n$ & $T_{vs}$(sec) & $T_{commn}$(sec) & $T_{exec}$(sec) & $T_{comp}$(sec) & $T_{Est}$(sec) & $T_{Rec}$(sec) \\ \hline
5      & 5   & 1.5            & 18               & 16              & 34              & 68.52          & 78             \\
5      & 10  & 3              & 9.88             & 8               & 17.88           & 53.88          & 72             \\
5      & 15  & 4.5            & 9.5              & 4               & 13.5            & 51             & 66             \\
5      & 20  & 6              & 9.3              & 2               & 11.4            & 50.4           & 54             \\
10     & 5   & 3              & 28.2             & 24              & 52.2            & 88.2           & 96             \\
10     & 10  & 6              & 7.74             & 12              & 19.74           & 58.74          & 72             \\
10     & 15  & 9              & 5.4              & 6               & 11.4            & 53.4           & 66             \\
10     & 20  & 12             & 3                & 3               & 6               & 51             & 54             \\
15     & 5   & 4.5            & 37.9             & 32              & 69.9            & 107.4          & 108            \\
15     & 10  & 9              & 8.3              & 16              & 24.6            & 63.6           & 78             \\
15     & 15  & 13.5           & 5.7              & 8               & 13.7            & 60.7           & 72             \\
15     & 20  & 18             & 2.4              & 4               & 6.4             & 57.4           & 60             \\
20     & 5   & 6              & 40.2             & 48              & 88.2            & 127.2          & 114            \\
20     & 10  & 12             & 12.2             & 24              & 36.2            & 81.4           & 84             \\
20     & 15  & 18             & 8.5              & 12              & 17.5            & 68.5           & 72             \\
20     & 20  & 24             & 6.2              & 6               & 12.2            & 68.52          & 60             \\ \hline
\end{tabular}
}
\label{table:calc}
\end{table}
Finally, combining the Equations \ref{eqn:brd} and \ref{eqn:comp3} in  Equation \ref{eqn:est1}, the estimated total
completion time for the target job is given as
\begin{align}\label{eqn:est2}
\begin{split}
 \mathit{T_{\mathit{Est}}}  =   \mathit{T_{\mathit{Init}}}  +    \mathit{T_{\mathit{prep}}} + n \times \mathit{iter} \times C  +   \mathit{iter} \times  B / n + \frac{A \times s}{n},
\end{split}
 \end{align} where $\mathit{n_{unit}}$ is the number of unit RDD tasks given in Equation \ref{eqn:s},
 $A$ = $\mathit{\frac{\mathit{cf}_{\mathit{commn}} \times T_{\mathit{commn}}^{\mathit{baseline}}}{\mathit{s_{\mathit{baseline}}}}}$,
   $B$ = $\sum_{\substack{k=1}}^{\mathit{n_{unit}}}
                                                                    \mathit{M_{a}^{k}}$, and $C$ = $\mathit{coeff} \times \mathit{T_{\mathit{vs}}^{\mathit{baseline}}}$. 
Table \ref{table:calc} shows the stepwise calculations for the length of the various phases in the estimated total
  completion time $\mathit{T_{\mathit{Est}}}$ for the MovieLensALS application, in standalone mode, with varying number of nodes $n$, and the number of iterations $\mathit{iter}$, on m1.large Ec2 instance.

\begin{figure*}[htpb]
\centering
\subfigure[Accuracy of Estimation for ALS With Varying Input DataSet Size in stand-alone mode with 20 worker nodes]{\includegraphics[width=0.3\linewidth,height=4.2cm]{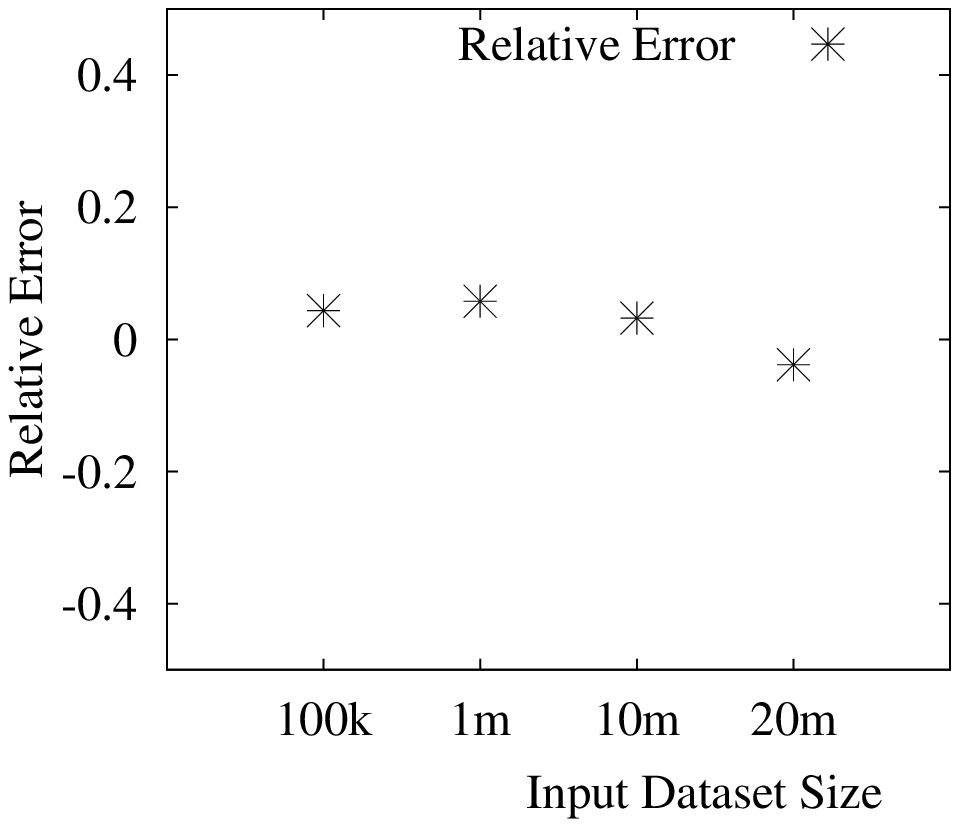}\label{fig:ALSStandSize}}\hfill
\subfigure[Accuracy of Estimation for  PageRank With Varying Input DataSet Size in stand-alone mode  20 worker nodes and 5 iterations]{\includegraphics[width=0.3\linewidth,height=4.2cm]{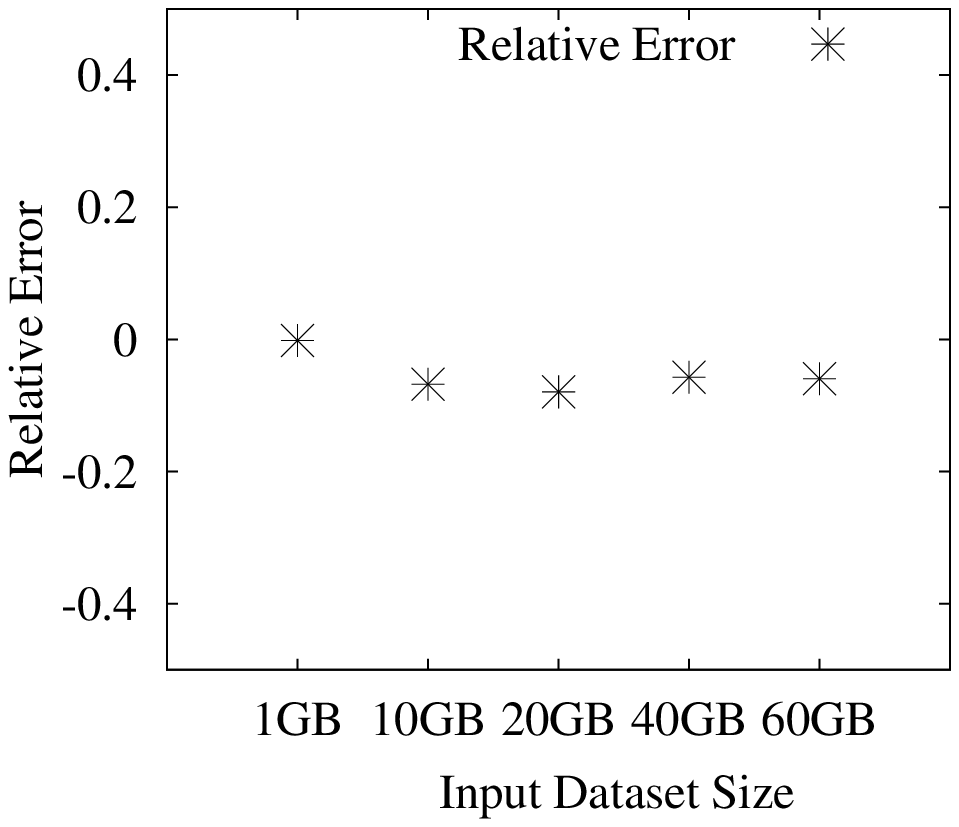}\label{fig:wordcountStandSize}}\hfill
\subfigure[Accuracy of Estimation for  WordCount With Varying Input DataSet Size in stand-alone mode with 20 worker nodes]{\includegraphics[width=0.3\linewidth,height=4.2cm]{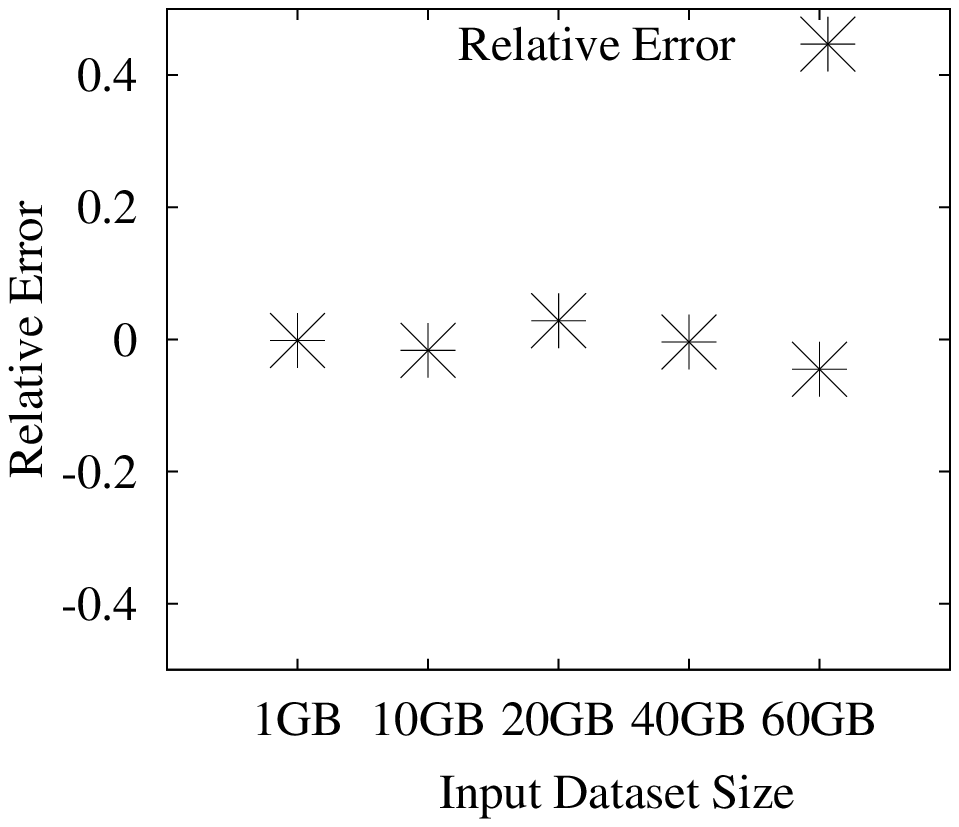}\label{fig:pagerankStandSize}}\hfill
\subfigure[Accuracy of Estimation for PageRank With Varying Number of Nodes in stand-alone mode]{ \includegraphics[width=0.3\linewidth,height=4.2cm]{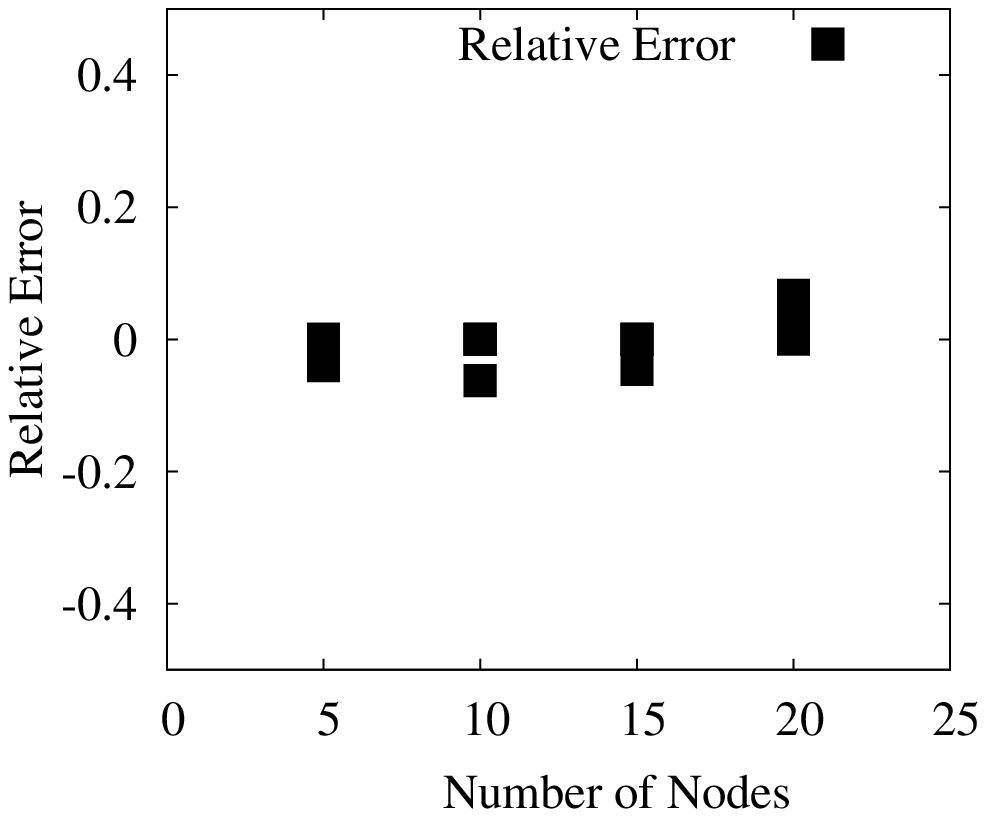}\label{fig:pagerankStand}}\hfill
\subfigure[Accuracy of Estimation for PageRank With Varying Number of Iterations in stand-alone mode]{ \includegraphics[width=0.3\linewidth,height=4.2cm]{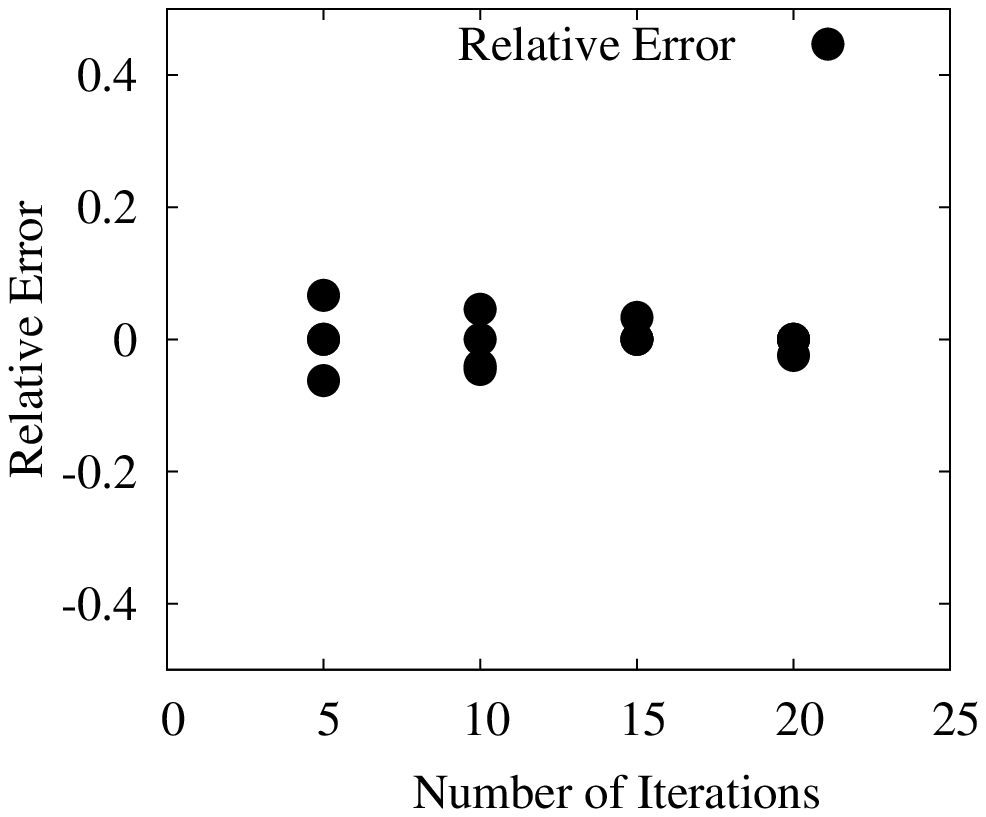}\label{fig:pagerankStanditer}}\hfill
\subfigure[Accuracy of Estimation for Logistic Regression With Varying Number of Nodes in stand-alone mode]{ \includegraphics[width=0.3\linewidth,height=4.2cm]{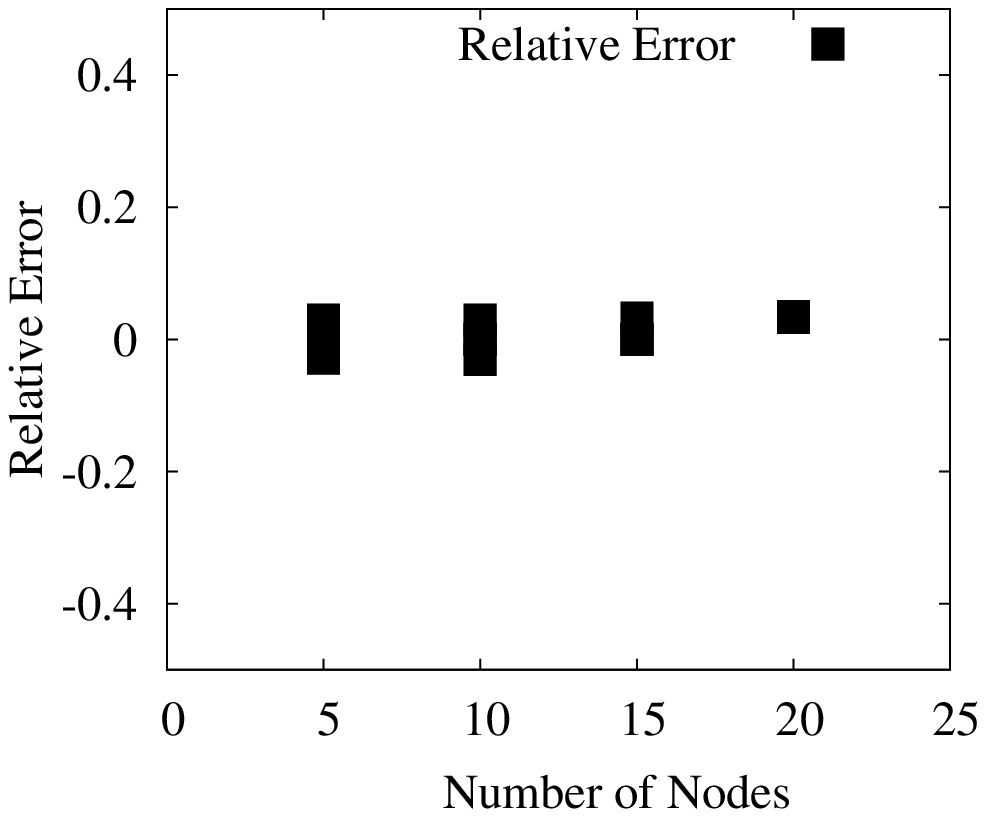}\label{fig:LRStand}}\hfill
\subfigure[Accuracy of Estimation for Logistic Regression With Varying Number of Iterations in stand-alone mode]{ \includegraphics[width=0.3\linewidth,height=4.2cm]{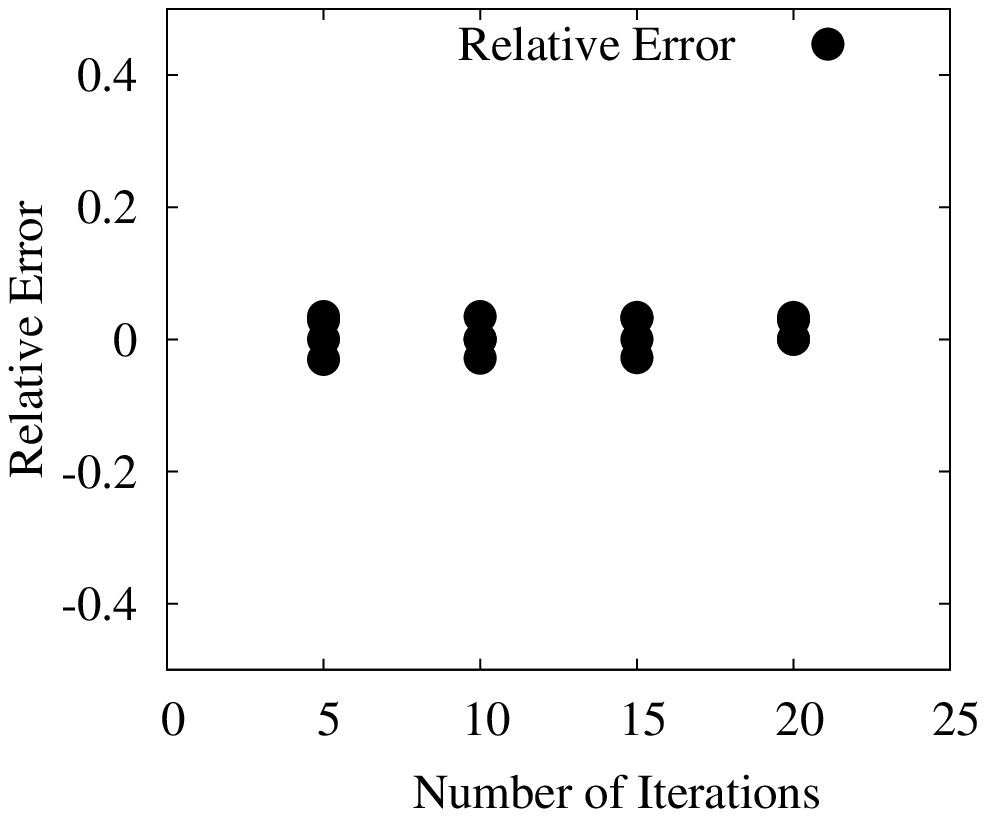}\label{fig:LRStanditer}}\hfill
\subfigure[Accuracy of Estimation for WordCount With Varying Number of Nodes in stand-alone mode]{ \includegraphics[width=0.3\linewidth,height=4.2cm]{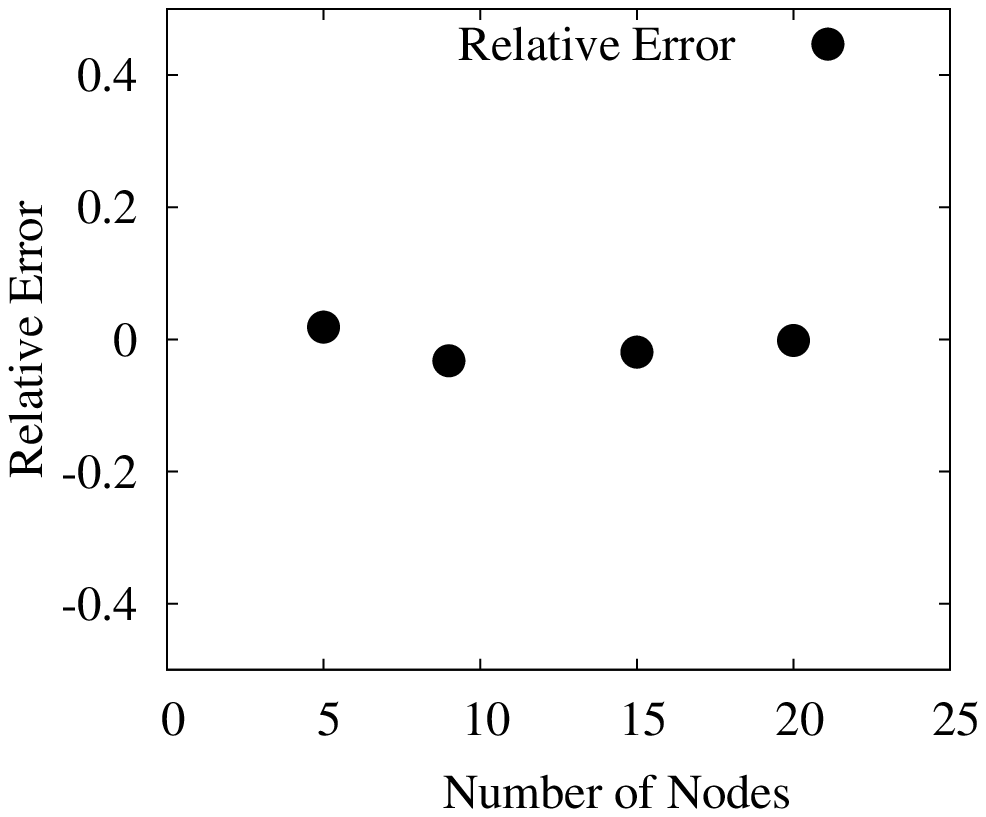}\label{fig:wordcountStand}}\hfill
\subfigure[Accuracy of Estimation for WordCount With Varying Number of Nodes in YARN mode]{ \includegraphics[width=0.3\linewidth,height=4.2cm]{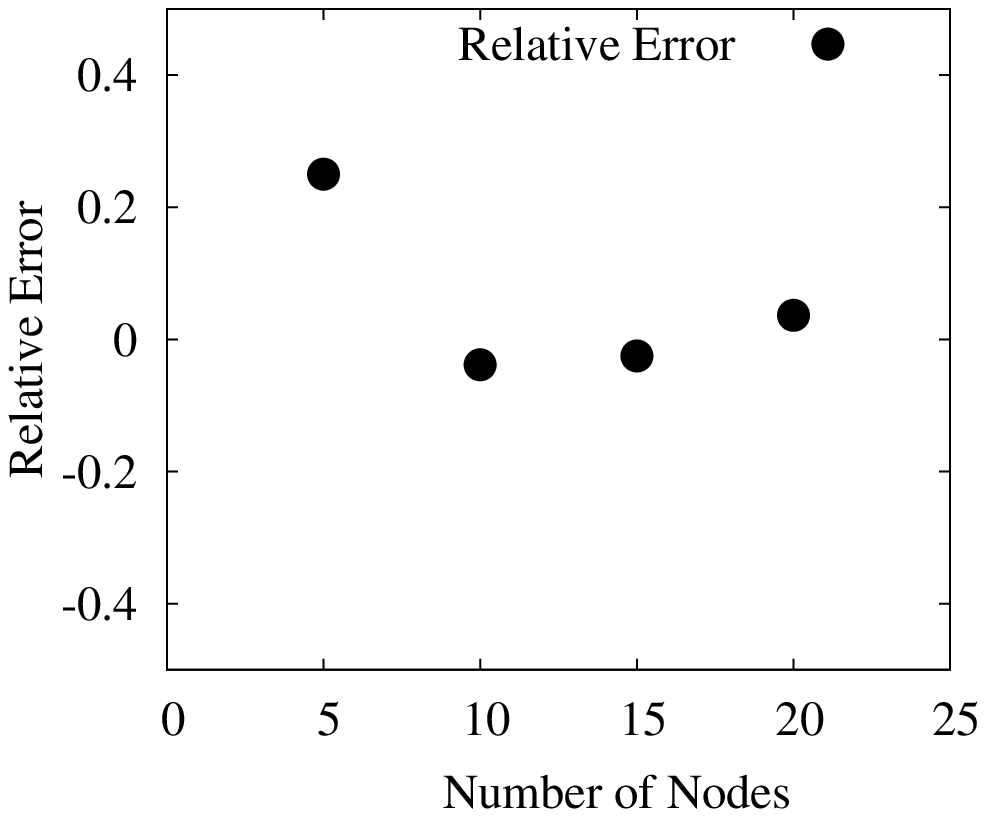}\label{fig:wordcountYarn}}\hfill
\caption{Accuracy Of Estimations against varying input dataset size, number of nodes and iterations}
\label{fig:RE1}
\end{figure*}

\begin{figure*}[htpb]
\subfigure[Accuracy of Estimation for PageRank With Varying Number of Nodes in YARN mode]{ \includegraphics[width=0.3\linewidth,height=4.2cm]{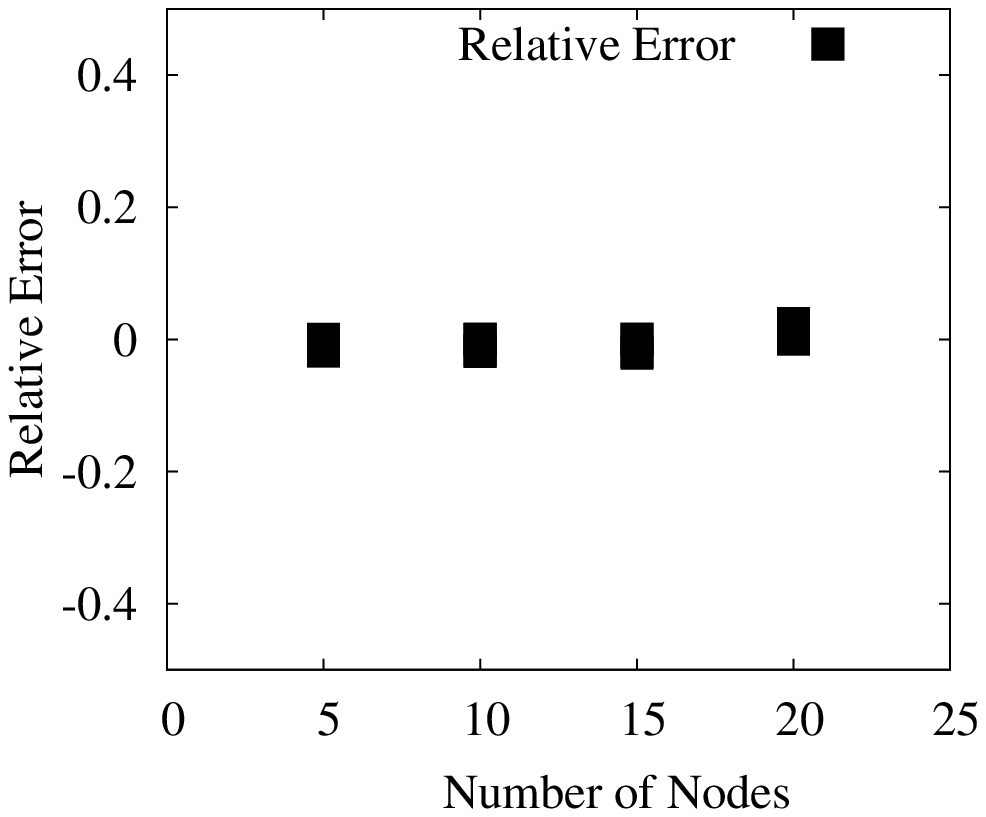}\label{fig:pagerankYarn}}\hfill
\subfigure[Accuracy of Estimation for PageRank With Varying Number of Iterations in YARN mode]{ \includegraphics[width=0.3\linewidth,height=4.2cm]{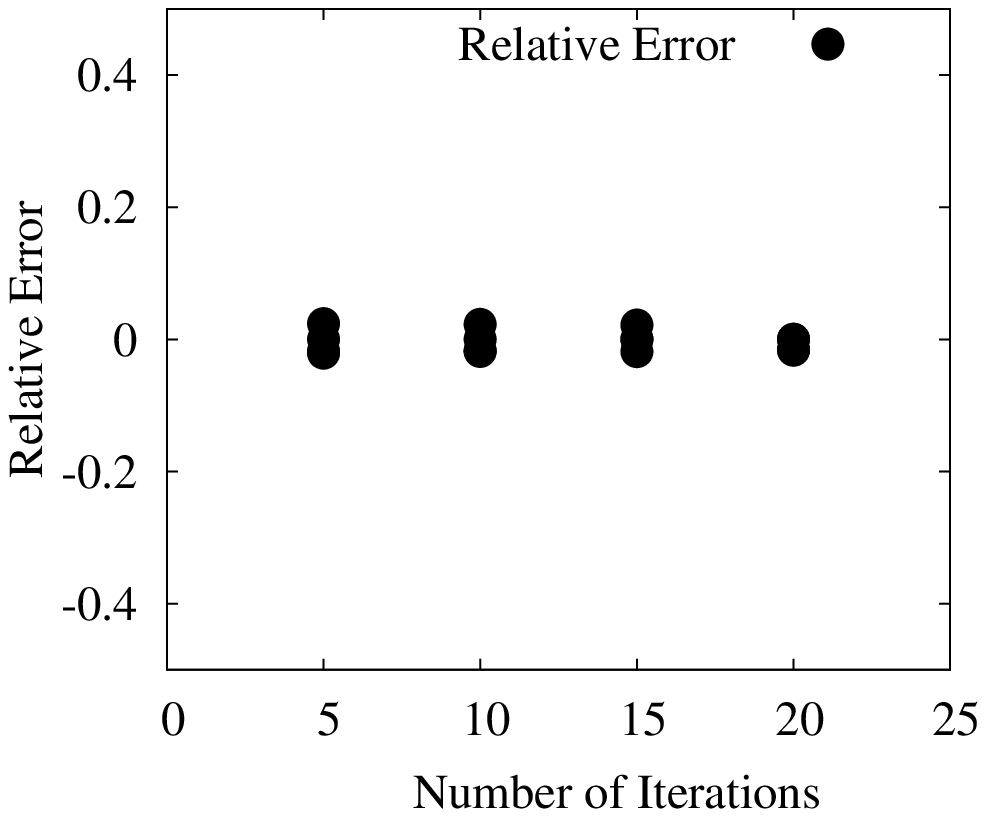}\label{fig:pagerankYarniter}}\hfill
\subfigure[Accuracy of Estimation for Logistic Regression With Varying Number of Nodes in YARN mode]{ \includegraphics[width=0.3\linewidth,height=4.2cm]{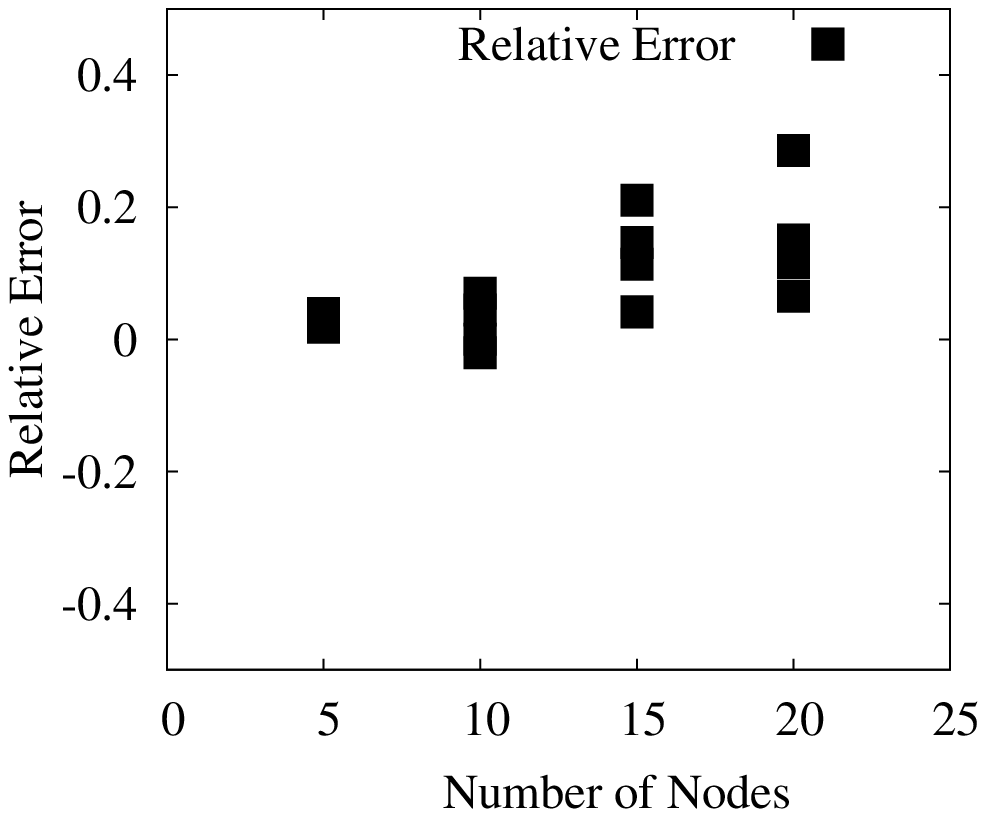}\label{fig:LRYarn}}\hfill
\subfigure[Accuracy of Estimation for Logistic Regression With Varying Number of Iterations in YARN mode]{ \includegraphics[width=0.3\linewidth,height=4.2cm]{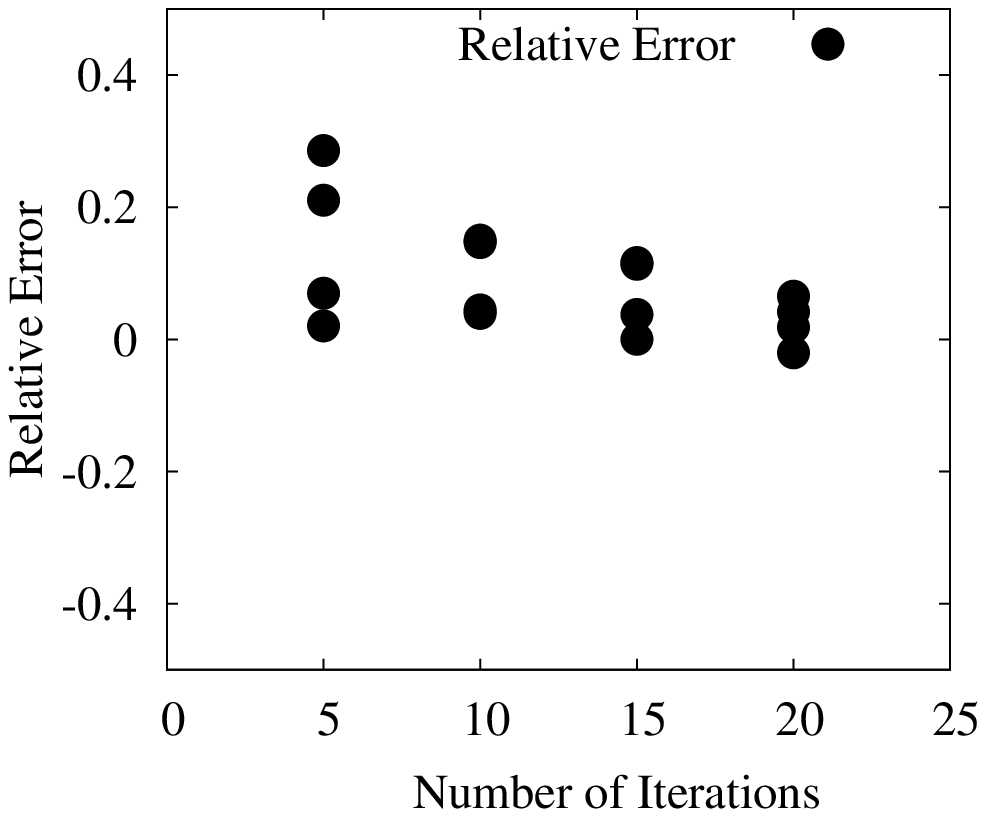}\label{fig:LRYarniter}}\hfill
\subfigure[Accuracy of Estimation for ALS With Varying Number of Nodes in stand-alone mode]{ \includegraphics[width=0.3\linewidth,height=4.2cm]{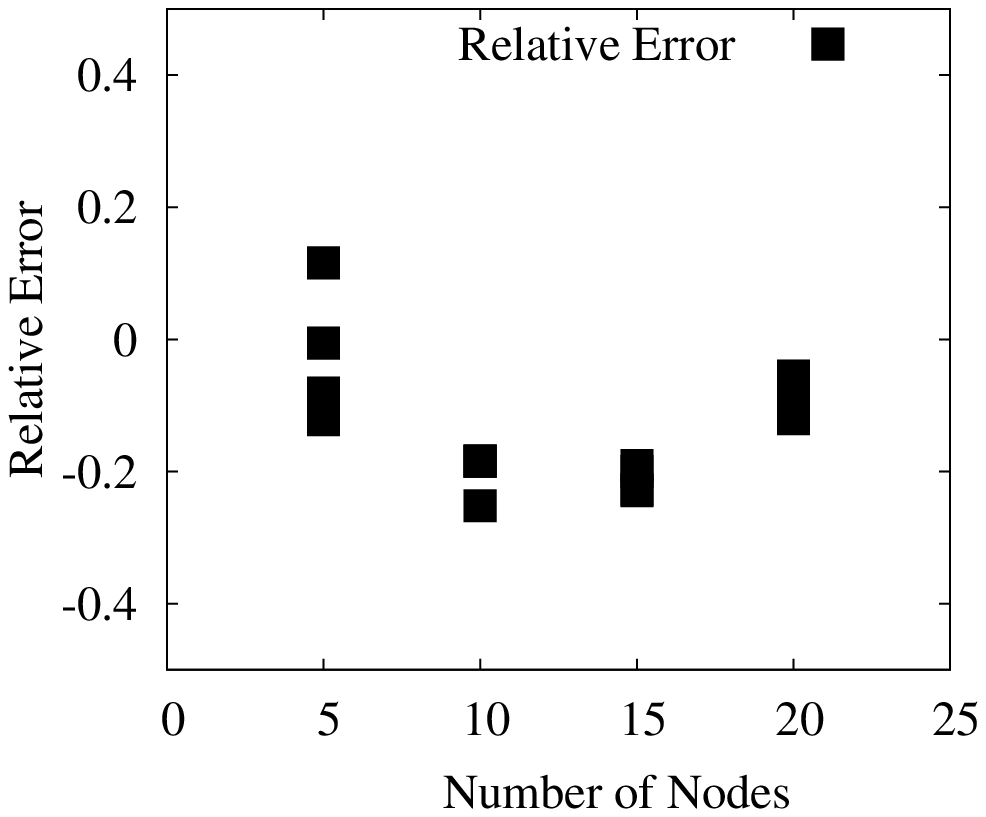}\label{fig:ALSStand}}\hfill
\subfigure[Accuracy of Estimation for ALS With Varying Number of Iterations in stand-alone mode]{ \includegraphics[width=0.3\linewidth,height=4.2cm]{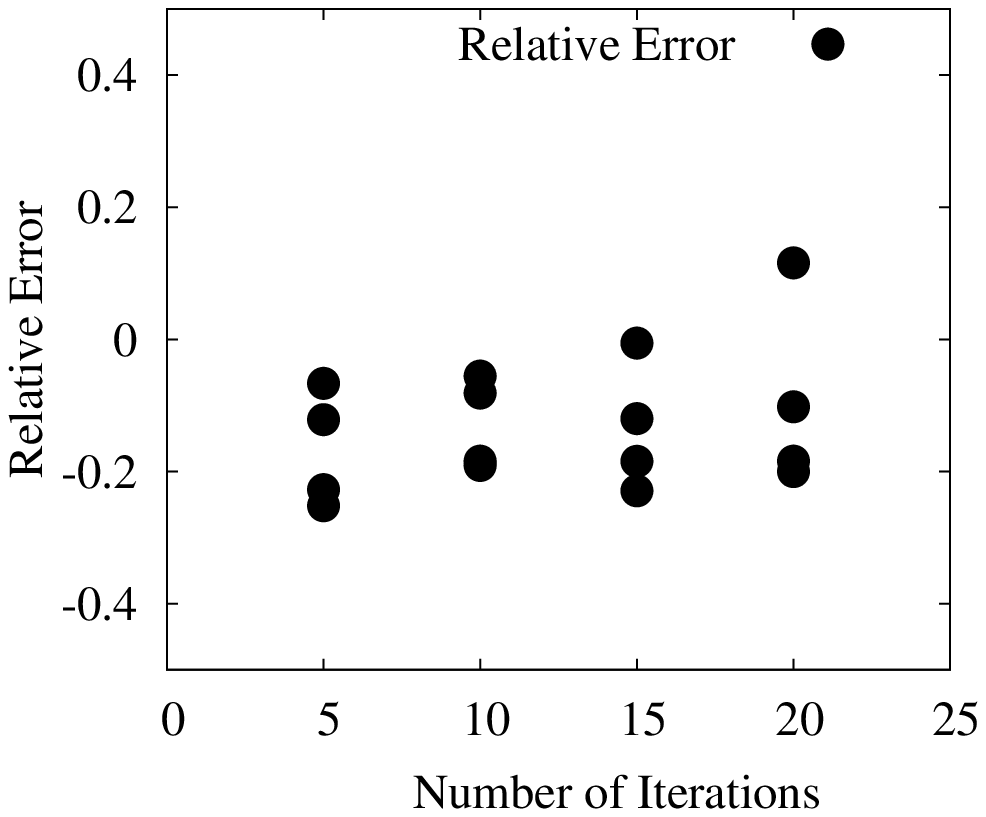}\label{fig:ALSStanditer}}\hfill
\subfigure[Accuracy of Estimation for ALS With Varying Number of Nodes in YARN mode]{ \includegraphics[width=0.3\linewidth,height=4.2cm]{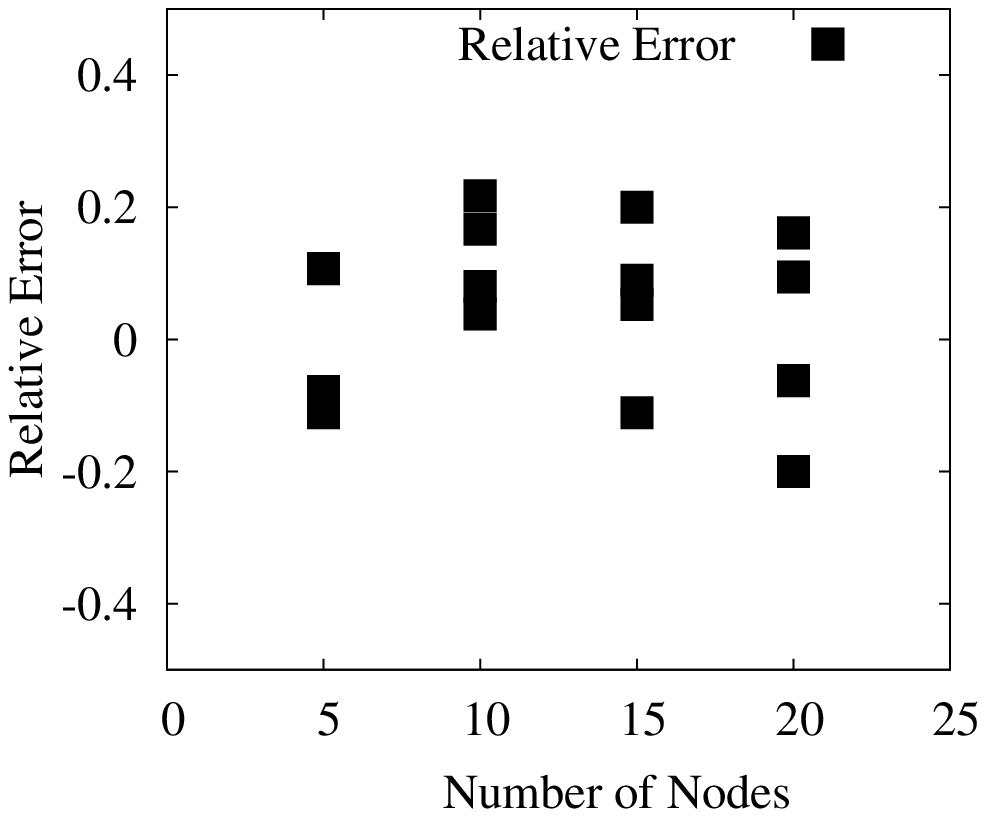}\label{fig:ALSYarn}}\hfill
\subfigure[Accuracy of Estimation for ALS With Varying Number of Iterations in YARN mode]{ \includegraphics[width=0.3\linewidth,height=4.2cm]{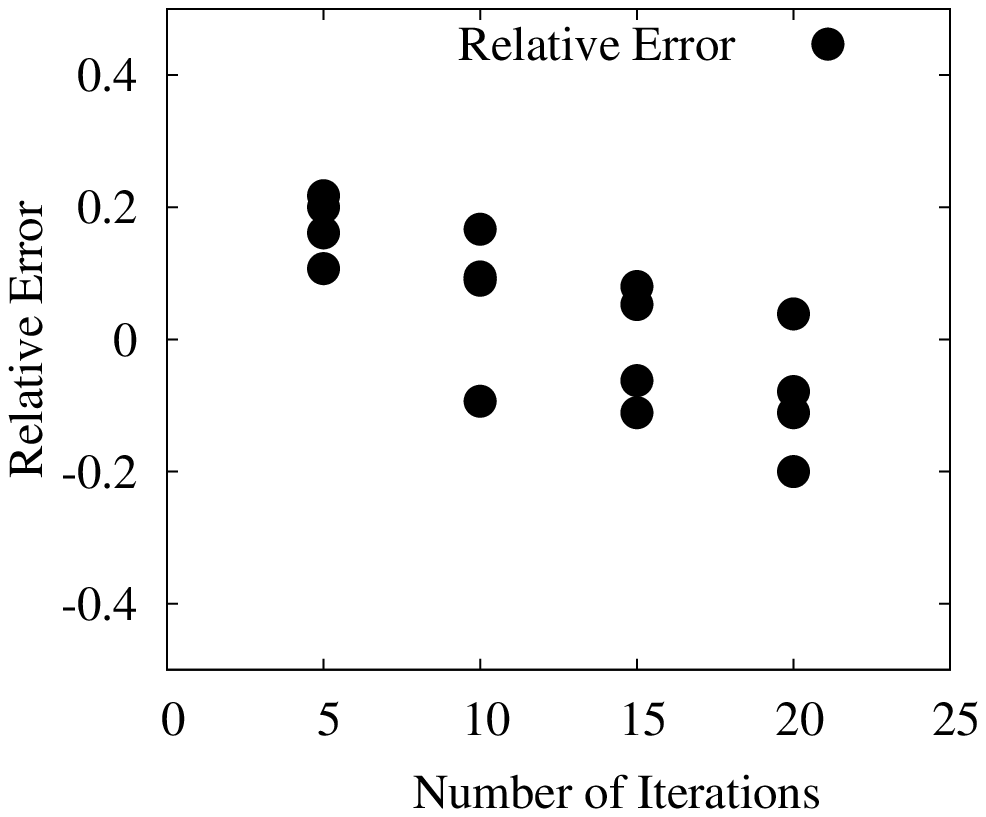}\label{fig:ALSYarniter}}\hfill
\subfigure[Mean Relative Error of OptEx Estimations]{
\scalebox{0.9}{
\begin{tabular}[b]{|c |c |c |c|} 
\hline\hline 
Application & Stand-alone & YARN \\ 
 \hline\hline
PageRank &  0.019 & 0.012 \\
Wordcount&  0.036 & 0.087 \\
LR &  0.020 & 0.086 \\
MovieLensALS & 0.094 & 0.126 \\ 
 \hline\hline
Mean & \multicolumn{2}{c|}{0.06} \\
Standard Deviation & \multicolumn{2}{c|}{0.04} \\
Variance & \multicolumn{2}{c|}{0.0016} \\
Range & \multicolumn{2}{c|}{0.012 - 0.13} \\
Majority Bound & \multicolumn{2}{c|}{0.02} \\
Confidence Interval & \multicolumn{2}{c|}{0.056-0.062} \\
\hline 
\end{tabular}\label{table:error}
}}
\caption{More Accuracy Results and the Observed Mean Relative Error Sattistics}
\label{fig:RE2}
\end{figure*}

\section{Estimation of cost optimal cluster composition}\label{sec:cost}

 OptEx models the completion time $T_{\mathit{Est}}$ (Equation
\ref{eqn:est2}) of a Spark job on a cluster comprising virtual machine instances provisioned
 from a cloud service provider, like Amazon (EC2), RackSpace, Microsoft, etc. The OptEx model is further used to estimate the cost optimal cluster composition for running a given job on
  virtual machine instances provided by any cloud provider, under
the job completion deadline specified in the SLO, while minimizing the service usage cost.
 Let the optimal cluster size be given as $n = \sum_{t=1}^m {\mathit{n_t}},$
 where $n_t$ is the number of virtual machine instances of type $t$ (depends on the instance offerings of the chosen cloud provider), and  $m$ is the total number of possible machine instance types.
 Let total service usage cost of running the given job on the cloud be denoted by $C$.
  Let $\mathit{c_{t}}$ be the hourly cost of each
 machine instance of type $t$ (depends on the current rates charged by the chosen cloud provider), and $\mathit{T_{\mathit{Est}}}$ be the estimated
 completion time of the given job (Equation \ref{eqn:est2}).
   Our objective is to determine the
 cost optimal cluster composition for finishing the given Spark job within an SLO deadline with minimum service usage cost.
  This goal can be mathematically stated as: optimize the objective function
\begin{equation}\label{eqn:obj}
 C = \sum_{t=1}^{m} \mathit{c_{t}} \times \mathit{n_{t}} \times \mathit{T_{\mathit{Est}}},
\end{equation}
and  obtain the cost optimal cluster composition, given as $\mathit{N_t}$ = \[ \Set{n_t} {1\le t\le m}{,} \] under
the constraint $\mathit{T_{\mathit{Est}}} < \mathit{SLO}$, where $\mathit{SLO}$ is the given deadline, and
 $T_{Est}$ is estimated using Equation \ref{eqn:est2}.
\par We optimize the above objective function (Equation \ref{eqn:obj}) 
 and determine an optimal cluster configuration given by $\mathit{N_t}$
, under the constraint $\mathit{T_{\mathit{Est}}} < \mathit{SLO}.$
The above constraint involving $\mathit{T_{\mathit{Est}}}$ is described as a convex  nonlinear function over $n$ (Equation
\ref{eqn:est2}), and is twice differentiable 
 with
respect to $n$, i.e., both first and second derivatives of $\mathit{T_{\mathit{Est}}}$ exist with respect to $n$.
The above optimization problem, for minimizing the cost $C$, under the nonlinear constraint $\mathit{T_{\mathit{Est}}} < \mathit{SLO}$ is solved using the Interior Point algorithm \cite{leader2004numerical}.
 The solution to the
above optimization problem enables: 1) estimating whether a
given job will finish under the deadline $\mathit{SLO}$, 2) optimal job
scheduling under the given deadline $\mathit{SLO}$, while minimizing
cost $C$, and 3) estimating optimal cluster
composition, given a cost budget $C$ and an $\mathit{SLO}$.  

\section{Implementation and Evaluation}\label{sec:evaluation}

\subsection{Experimental Setup}\label{sec:setup}
 The experimental setup consists of Apache Spark version 1.2.1, built-in within the Cloudera Express 5.3.1 package, on a
  cluster of m1.xlarge Amazon EC2 machine instances, each comprising 8 cores, 15 GB of RAM, and 10 GB EBS
, and
  running RedHat Enterprise Linux version 6.
   We use HDFS as the backend for storing and processing the input dataset. The underlying Hadoop cluster has a namenode, a secondary namenode, and
5 datanodes under a replication factor of 3. 
 \subsection{Experimental Procedure}\label{sec:procedure}
  We use the Interior point algorithm \cite{leader2004numerical} from the Optimization toolbox of the Matlab version 2013b 
 for solving the given non-linear convex optimization problem (Section \ref{sec:cost}), and estimating the cost optimal cluster composition. 
We use the default FIFO scheduler. The average scheduler delay is
 4 ms, and can be neglected relative to the other components of the execution time.
     The input workload for
the MovieLensALS application is the 10-M MovieLens dataset obtained from grouplens.org \cite{MOVIELENS-DATA}.
 PageRank is evaluated with the social network dataset for LiveJournal \cite{snapnets}, an online community comprising roughly 10 million members. 
  The LiveJournal dataset has 4847571 nodes and 68993773 edges. 
 The input workload for the Wordcount application are the Wikipedia dumps obtained from SNAP \cite{snapnets}.

\subsection{Technique for Generating Job Profiles}\label{sec:tech}
 For computing the job profile of a target given
 job $j$ with respect to a virtual machine instance type $t$, we run the representative job $a$ (Section \ref{sec:rep}) for the
 application category corresponding to the target job $j$ (Section \ref{sec:cat}), 
  on a cloud instance of type $t$ in stand-alone mode with a benchmark workload \cite{Pavlo:2009:CAL:1559845.1559865}. 
We estimate the components of the job profile from the snapshots of the execution flow of the representative
   job obtained using YourKit Java Profiler \cite{YKfJ07}. 
  To minimize overhead, YourKit is run in the sampling mode. 

\subsection{Accuracy of the Estimations Using OptEx}\label{sec:eval}

    Being the first work in modelling Spark jobs, OptEx has no prior baseline to compare with. However,
    we demonstrate (see Figures \ref{fig:RE1} and \ref{fig:RE2}) that OptEx provides accurate (i.e., average relative error 0.06) estimations of the job completion time against variations
    in all the input parameters of the model (i.e, against increasing size of dataset, number of nodes, and the
    number of iterations), and on applications of different categories. 
 From the estimated completion time $T_{\mathit{Est}}$ and the recorded (i.e., observed) completion time
 $T_{\mathit{Rec}}$, we compute the relative error
 $\mathit{RE} = (T_{\mathit{Est}} - T_{\mathit{Rec}})/T_{\mathit{Rec}}$.
 The Figures \ref{fig:ALSStandSize}, \ref{fig:wordcountStandSize}, and \ref{fig:pagerankStandSize} illustrate
 the variations in 
 the relative error $\mathit{RE}$, for the MovieLensALS, Wordcount, and PageRank
   applications, with increasing size $s$ of the input dataset.
    \par 
The figures \ref{fig:wordcountStand}, \ref{fig:pagerankStand}, \ref{fig:LRStand}, and \ref{fig:ALSStand} 
 illustrate the variations in the relative error  $\mathit{RE}$, 
  for MovieLensALS, Wordcount, PageRank, and Logistic Regression, against
 varying cluster size $n$, in the
 stand-alone mode. The figures \ref{fig:pagerankStanditer}, \ref{fig:LRStanditer}, and \ref{fig:ALSStanditer} 
 illustrate the variations in the relative error  $\mathit{RE}$ 
  for the same applications, against varying
 number of iterations $\mathit{iter}$, in the
 stand-alone mode. Figures \ref{fig:wordcountYarn}, \ref{fig:pagerankYarn}, \ref{fig:LRYarn}, and \ref{fig:ALSYarn} 
  illustrate the variations in  the relative error $\mathit{RE}$ 
 for the same applications with varying $n$ in the  YARN mode.  Figures \ref{fig:pagerankYarniter},
 \ref{fig:LRYarniter}, and \ref{fig:ALSYarniter} 
  illustrate the variations in the relative error $\mathit{RE}$ 
 for the same applications against varying $\mathit{iter}$, in
 YARN mode.
 \begin{table*}[!ht]
\caption{Optimal Scheduling With Estimated Optimal Cluster Size Under Varying SLO} 
\scalebox{0.9}{
\resizebox{\textwidth}{!}{\begin{tabular}{|l|l|l|c|l|l|l|l|l|l|l|l|l|l|l}
\hline
\multirow{3}{*}{SLO(sec)} & \multicolumn{1}{c|}{\multirow{3}{*}{Mode}} & \multicolumn{1}{c|}{\multirow{3}{*}{App}} & \multicolumn{12}{c|}{Iterations}                                                                                                                                                \\ \cline{4-15}
                     & \multicolumn{1}{c|}{}                      & \multicolumn{1}{c|}{}                             & \multicolumn{3}{c|}{5}                          & \multicolumn{3}{c|}{10}                        & \multicolumn{3}{c|}{15}    & \multicolumn{3}{c|}{20}                         \\ \cline{4-15}
                     & \multicolumn{1}{c|}{}                      & \multicolumn{1}{c|}{}                             & \multicolumn{1}{l|}{n}  & $T_{Est}$(sec) & $T_{Rec}$(sec) & n                      & $T_{Est}$(sec) & $T_{Rec}$(sec) & n  & $T_{Est}$(sec) & $T_{Rec}$(sec) & n  & $T_{Est}$(sec) & \multicolumn{1}{l|}{$T_{Rec}$(sec)} \\ \hline
200                  & Standalone                                 & \multicolumn{1}{c|}{ALS}                  & 3                       & 103.2     & 107       & \multicolumn{1}{c|}{5} & 88        & 93        & 6  & 89.67     & 92        & 7  & 88.84     & \multicolumn{1}{l|}{95}        \\ \hline
240                  & YARN                                       & ALS                                       & \multicolumn{1}{l|}{2}  & 240       & 250       & 3                      & 228.44    & 235       & 4  & 211.5     & 215       & 4  & 240       & \multicolumn{1}{l|}{237}       \\ \hline
350                  & Standalone                                 & \multicolumn{1}{c|}{Wordcount}                & 1                       & 332.8     & 345       & \multicolumn{9}{l|}{\multirow{2}{*}{}}                                                                                        \\ \cline{1-6}
800                  & YARN                                       & Wordcount                                     & 1                       & 794.8     & 821       & \multicolumn{9}{l|}{}                                                                                                         \\ \hline
150                  & Standalone                                 & ALS                                       & \multicolumn{1}{l|}{3}  & 100       & 97        & 4                      & 99        & 97        & 4  & 140.5     & 143       & 6  & 103.22    & \multicolumn{1}{l|}{113}       \\ \hline
200                  & YARN                                       & ALS                                       & \multicolumn{1}{l|}{3}  & 174.22    & 215       & 4                      & 181       & 185       & 5  & 178.56    & 186       & 5  & 198       & \multicolumn{1}{l|}{195}       \\ \hline
330                  & Standalone                                 & Wordcount                                     & 2                       & 321.69    & 325       & \multicolumn{9}{l|}{\multirow{2}{*}{}}                                                                                        \\ \cline{1-6}
790                  & YARN                                       & Wordcount                                     & 2                       & 783.69    & 781       & \multicolumn{9}{l|}{}                                                                                                         \\ \hline
100                  & Standalone                                 & ALS                                       & \multicolumn{1}{l|}{4}  & 79.5      & 75        & 6                      & 76.11     & 78        & 6  & 89.67     & 91        & 7  & 88.84     & \multicolumn{1}{l|}{94}        \\ \hline
160                  & YARN                                       & ALS                                       & 4                       & 150.5     & 157       & 5                      & 159       & 158       & 6  & 160       & 155       & 7  & 159.84    & \multicolumn{1}{l|}{158}       \\ \hline
325                  & Standalone                                 & Wordcount                                     & 2                       & 321.69    & 323       & \multicolumn{9}{l|}{\multirow{2}{*}{}}                                                                                        \\ \cline{1-6}
785                  & YARN                                       & Wordcount                                     & 2                       & 783.69    & 782       & \multicolumn{9}{l|}{}                                                                                                         \\ \hline
75                   & Standalone                                 & ALS                                       & \multicolumn{1}{l|}{5}  & 68.52     & 65        & 7                      & 75     & 67        & 8  & 71.88     & 72        & 9  & 73.1      & \multicolumn{1}{l|}{71}        \\ \hline
140                  & YARN                                       & ALS                                       & \multicolumn{1}{l|}{5}  & 139.52    & 138       & 7                      & 139.92    & 135       & 9  & 138.07    & 133       & 9  & 139.1     & \multicolumn{1}{l|}{135}       \\ \hline
320                  & Standalone                                 & Wordcount                                     & 3                       & 319.64    & 315       & \multicolumn{9}{l|}{\multirow{2}{*}{}}                                                                                        \\ \cline{1-6}
783                  & YARN                                       & Wordcount                                     & 3                       & 781.64    & 754       & \multicolumn{9}{l|}{}                                                                                                         \\ \hline
60                   & Standalone                                 & ALS                                       & \multicolumn{1}{l|}{7}  & 58.96     & 56        & 10                     & 58.76     & 57        & 11 & 61.1      & 55        & 12 & 62.56     & \multicolumn{1}{l|}{58}        \\ \hline
121                  & YARN                                       & ALS                                       & \multicolumn{1}{l|}{22} & 121.05    & 118       & 31                     & 121.02    & 115       & 39 & 120.97    & 112       & 42 & 121.1     & \multicolumn{1}{l|}{113}       \\ \hline
319                  & Standalone                                 & Wordcount                                     & 4                       & 318.93    & 332       & \multicolumn{9}{l|}{\multirow{2}{*}{}}                                                                                        \\ \cline{1-6}
781                  & YARN                                       & Wordcount                                     & 4                       & 780.93    &  772         & \multicolumn{9}{l|}{}                                                                                                         \\ \hline
\end{tabular}}}
\label{table:minimize}
\end{table*}
\begin{table*}[!ht]
\scalebox{0.9}{
\resizebox{\textwidth}{!}{\begin{tabular}{|c|l|l|l|c|l|l|l|l|c|l|}
\hline
Category            & \multicolumn{4}{c|}{Standalone}                                                         & \multicolumn{4}{c|}{YARN}                                                           \\ \hline
\multicolumn{1}{|l|}{} & Standard Deviation & Variance & Mean   & \multicolumn{1}{l|}{Confidence} & Standard Deviation & Variance & Mean   & \multicolumn{1}{l|}{Confidence} \\ \hline
MLlib            & 0.373              & 0.139    & 1.106  & 0.183           & 0.702              & 0.493    & 2.2790 & 0.344                               \\ \hline
Spark Streaming          & 0.265              & 0.07     & 5.7    & 2.794                 & 2.85       & 8.15     & 13.275 & 1.4                                  \\ \hline
Spark SQL                 & 1.96               & 3.85     & 32.63  & 0.96                                       & 3.263              & 10.65    & 48.88  & 0.087                                 \\ \hline
GraphX               & 8.348              & 69.696   & 26.687 & 4.38                                      & 5.86               & 34.4     & 53     & 2.874                                 \\ \hline
\end{tabular}}}
\caption{Confidence Interval of Estimation With Varying Choice of Representative Jobs for Each Job Category}
\label{table:confidence} 
\end{table*}
  We evaluate the OptEx model with the
 mean relative error metric \cite{leader2004numerical}
  $\delta$ = $\frac {\sum_{j=1}^{k} \abs{T_{\mathit{Est}} - T_{\mathit{Rec}}}/T_{\mathit{Rec}}} {k}$, where $k$ is the total number of jobs submitted. 
   The absolute differences between
   $T_{\mathit{Est}}$ and $T_{\mathit{Rec}}$ eliminate the signs in the error, and gives the magnitudes of the errors. The error values $\delta$ 
 are
given in the Table \ref{table:error}. The average $\delta$ score for all the cases is 0.06, i.e., 6\%. 
$\newcommand{\E}{\mathrm{E}}$

\subsection{Analysis of the Results}\label{sec:discussion}
  The magnitude of the relative error  $\mathit{RE}$ for the experiments with Wordcount, PageRank, and Logistic Regression
 applications, representing the Streaming and GraphX categories (Section \ref{sec:cat}), in stand-alone mode, is strictly within 0-0.06, (Figures \ref{fig:ALSStandSize} through 
  \ref{fig:pagerankYarniter}), bounded by a 95\% confidence interval  of 0.056-0.062 (Table \ref{table:error}), except for one observation in Figure \ref{fig:wordcountYarn}.   
 The experiments with increasing dataset size yield relative error of magnitude between 0.007 to 0.05 (Figures \ref{fig:ALSStandSize}, \ref{fig:wordcountStandSize}, \ref{fig:pagerankStandSize}). 
         \par The estimated Spark job execution time $\mathit{T_{\mathit{Est}}}$ comprises two components \emph{$X_1$} and \emph{$X_2$}, where $X_1$ = $\mathit{T_{\mathit{Init}}}  +    \mathit{T_{\mathit{prep}}}$ and $X_2$ = $n \times \mathit{iter} \times C  +   \mathit{iter} \times  B / n + \frac{A \times s}{n}$ (Equation \ref{eqn:est2}). The first component $X_1$ is independent of variations in the values of the input variables. The second component $X_2$ comprises the last three phases of the Spark job execution (Section \ref{sec:model}), each phase varying differently with respect to the input variables $n$, $\mathit{iter}$, and $s$ (Section \ref{sec:var}). Hence, $X_2$ accounts for the observed random variations in the relative error, 
      with respect to variations in the input variables (Figures \ref{fig:RE1} and \ref{fig:RE2}).
         \par The execution phases in $X_2$ encompass the execution of the job stages, comprising unit RDD tasks, on the worker nodes (Section \ref{sec:formulation}). The execution of the job stages on the workers is inherently non-deterministic (unpredictable) in nature, due to the dependency on various components of the Spark cluster, like the driver, the cluster manager, the workers, etc., \cite{Zaharia:2012:RDD:2228298.2228301}. The job stages may get unpredictably delayed, i.e., can fail and get retried by the master repeatedly, due to various factors like momentary unavailability of required resources, delays in allocation of resources by the master, communication delays among the workers, etc., \cite{Zaharia:2012:RDD:2228298.2228301}. The above unpredictable delays in the job stages, however small, can cause the observed values of $X_2$  to deviate randomly from the estimated values of $X_2$, while $X_1$  stays constant (Section \ref{sec:formulation}). This, in turn, causes the overall observed completion time $\mathit{T_{\mathit{Rec}}}$ of the job, to vary unpredictably with respect to the estimated job completion time $\mathit{T_{\mathit{Est}}}$, estimated from $X_1$  and $X_2$  (Section \ref{sec:model}). This causes the observed random variations in the values of the relative error (i.e., $RE$ = $T_{\mathit{Est}} - T_{\mathit{Rec}}$), though still bounded by the confidence interval of 0.056-0.062 (Figures \ref{fig:RE1} and \ref{fig:RE2}). 
      \par The relative error increases slightly with increasing number of nodes (Figures \ref{fig:wordcountStand}, \ref{fig:pagerankStand}, \ref{fig:LRStand}, \ref{fig:wordcountYarn}, \ref{fig:pagerankYarn}, \ref{fig:LRYarn}, \ref{fig:ALSStandSize}, \ref{fig:ALSStand}, and \ref{fig:ALSYarn}). Worker nodes increasing in number augments the chances of unpredictable failures of the job stages due to dependency on communication between a larger number of nodes, causing unpredictable variations in the component $X_2$ of the overall job completion time $\mathit{T_{\mathit{Est}}}$. This, in turn, causes the observed job completion time $\mathit{T_{\mathit{Rec}}}$ to deviate more unpredictably from the estimated completion time  $\mathit{T_{\mathit{Est}}}$, estimated from $X_1$ and $X_2$. The result is greater variation in the relative error (i.e., $RE$ = $T_{\mathit{Est}} - T_{\mathit{Rec}}$) with increasing number of nodes.
       Our goal is to
  provide correct estimations for SLO-driven user-facing applications. Few user-facing applications, that work under an SLO deadline, will require more than 50 nodes \cite{spark:usecase1234, spark:usecase5678}. OptEx can provide estimations for typical SLO-driven user-facing applications with a relative error close to 0 (Figures \ref{fig:RE1} and \ref{fig:RE2}). Applications that do not meet this criteria are batch processing applications, like
  bioinformatics, genomics, data analytics applications, etc., which typically do not work under a deadline \cite{spark:usecase9012}.
      \par For experiments run in YARN mode, the variations in the observed relative error, with respect to the variations in the input variables, are noticeably larger than the experiments run in stand-alone mode (Figures \ref{fig:pagerankYarn}, and \ref{fig:LRYarn}). In YARN mode, the submitted jobs are additionally dependent on the YARN resource manager to allocate resources, and to execute the jobs on the worker \cite{HadoopYARN}. Hence, the chances of unpredictable delays in the intermediate stages of a job are greater in YARN mode due to additional communication between the YARN resource manager and the Spark master \cite{HadoopYARN}. Thus, the chances of observing randomness in the relative error is greater for applications run in the YARN mode, 
   though the magnitude of the average error is 0.04. 
    Further, the relative error, for YARN mode, is even closer to 0 for  applications with number of iterations larger than 10, representing production level use cases \cite{spark:usecase1234, spark:usecase5678} (Figures \ref{fig:pagerankYarn}, \ref{fig:LRYarn}, and \ref{fig:ALSYarn}).
\par OptEx cannot account for the non-deterministic delays in communicating the intermediate RDD objects among the worker nodes during the execution of an iterative Spark job on the workers \cite{Zaharia:2012:RDD:2228298.2228301}. The above delays result in deviations in the observed length of the job stages, comprising the component $X_2$, from the estimated completion time \cite{Zaharia:2012:RDD:2228298.2228301}. For experiments with large number of iterations, the job stages in the initial iterations cache the intermediate RDD objects locally in the worker nodes, resulting in a decrease in the time spent in communicating the RDD objects among the workers during the later iterations \cite{Zaharia:2012:RDD:2228298.2228301}.  This results in a decrease in the deviation in the observed length of the job stages comprising $X_2$ from the estimated lengths of the stages. This, in turn, reduces the deviations in the overall observed completion time $\mathit{T_{\mathit{Rec}}}$ with respect to the estimated overall completion time $\mathit{T_{\mathit{Est}}}$, estimated from $X_1$ and $X_2$. 
 Indeed, with increasing number of iterations, a decreasing trend
  is observed in the relative error (Figures \ref{fig:RE1} and \ref{fig:RE2}). 
    So, we believe that OptEx can provide more accurate estimations with typical production level use cases, which typically involve number of iterations larger than 10 \cite{spark:usecase1234, spark:usecase5678}.
\subsection{Optimal Scheduling and Project Planning using OptEx}\label{sec:sched}
\begin{table}[!ht]
\caption{Optimal Scheduling With Estimated Optimal Cluster Size Under Given Cost Budget} 
\scalebox{0.9}{
\begin{tabular}{|l|l|l|c|l|l|}
\hline
Budget(\$) & Mode       & App                        & $n$ & \begin{tabular}[x]{@{}l@{}}$T_{Est}$\\(sec)\end{tabular} & \begin{tabular}[x]{@{}l@{}}$T_{Rec}$\\(sec)\end{tabular} \\ \hline
0.3    & Standalone & \multicolumn{1}{c|}{ALS}   & 53  & 49.17     & 48        \\ \hline
0.8    & YARN       & ALS                        & 58  & 120.15    & 115       \\ \hline
1      & Standalone & \multicolumn{1}{c|}{Wordcount} & 27  & 318.02    & 321       \\ \hline
1.5    & YARN       & Wordcount                      & 16  & 780.05    & 775       \\ \hline
0.2    & Standalone & ALS                        & 35  & 49.4      & 50        \\ \hline
0.5    & YARN       & ALS                        & 36  & 120.01    & 119       \\ \hline
0.8    & Standalone & Wordcount                      & 22  & 318.03    & 321       \\ \hline
1.2    & YARN       & Wordcount                      & 13  & 780.09    & 780       \\ \hline
0.15   & Standalone & ALS                        & 26  & 49.72     & 50        \\ \hline
0.4    & YARN       & ALS                        & 29  & 120.58    & 117       \\ \hline
0.6    & Standalone & Wordcount                      & 16  & 318.06    & 311       \\ \hline
1      & YARN       & Wordcount                      & 11  & 780.12    & 757       \\ \hline
0.1    & Standalone & ALS                        & 17  & 50.69     & 52        \\ \hline
0.3    & YARN       & ALS                        & 21  & 121.1     & 125       \\ \hline
0.4    & Standalone & Wordcount                      & 11  & 318.12    & 315       \\ \hline
0.8    & YARN       & Wordcount                      & 8   & 780.24    & 780       \\ \hline
0.08   & Standalone & ALS                        & 13  & 51.89     & 50        \\ \hline
0.2    & YARN       & ALS                        & 14  & 122.49    & 120       \\ \hline
0.1    & Standalone & Wordcount                      & 5   & 318.6     & 310       \\ \hline
0.5    & YARN       & Wordcount                      & 4   & 780.94    &  780         \\ \hline
\end{tabular}
}
\label{table:minimize1}
\end{table}
 Again, being the first work in modeling the execution time of Spark jobs, OptEx has no prior results to compare directly with.
  The closest work is Elastisizer \cite{Herodotou:2011:NOS:2038916.2038934}, which predicts optimal cluster composition for Hadoop, but does not address Spark.
   Moreover, Elastisizer over predicts, on an average by 20.1\% and worst case 58.6\% \cite{Herodotou:2011:NOS:2038916.2038934}. Since OptEx uses a closed-form to estimate the completion time, it does not suffer from over-prediction.
    Table \ref{table:minimize} demonstrates the effectiveness of the constrained optimization techniques of OptEx (Section \ref{sec:cost}) in designing optimal scheduling strategies.
 For each application (refer to the 3\textsuperscript{rd} column of Table \ref{table:minimize}) running in Standalone or YARN mode (the
 2\textsuperscript{nd} column), Table  \ref{table:minimize} gives the cost optimal cluster composition
 (the optimal cluster size $n$ is given in the 4\textsuperscript{th}, 7\textsuperscript{th}, 10\textsuperscript{th} and
 13\textsuperscript{th} columns against 5, 10, 15, and 20 iterations, respectively) for executing a given job under given SLO deadlines (the 1\textsuperscript{st} column), while minimizing
  the cost of usage of the virtual machine instances. The 5\textsuperscript{th},  8\textsuperscript{th}, 11\textsuperscript{th}, and
 14\textsuperscript{th} columns of Table \ref{table:minimize}
  give the completion times $T_{Est}$ estimated using OptEx, for 5, 10, 15, and 20 iterations, respectively. The
  6\textsuperscript{th},  9\textsuperscript{th}, 12\textsuperscript{th}, and
 15\textsuperscript{th} columns of Table \ref{table:minimize} give the recorded completion times $T_{Rec}$ with the estimated cluster composition.  
 \par Following \cite{Verma:2011:AAR:1998582.1998637}, we propose a statistic $\mathcal{S}$ to measure the effectiveness of
OptEx in estimating whether a given job will satisfy the SLO deadline, while minimizing the cost. $\mathcal{S}$ gives the percentage of cases
which did not violate the SLO deadline, in the experiments recorded in the Table
\ref{table:minimize}. $\mathcal{S}$ evaluates to approximately 98\%, which proves that OptEx is, in
fact, very effective for scheduling Spark jobs on the cloud, while minimizing the service usage cost. 
     \par Table \ref{table:minimize1} demonstrates that OptEx can be used in project planning for optimal cluster
 provisioning under given budget, while optimizing job execution times. Table \ref{table:minimize1} records the optimal cluster size (the 4\textsuperscript{th} column)
 required to run a given application (the 3\textsuperscript{rd} column) in Standalone or YARN mode (the
 2\textsuperscript{nd} column) estimated using Equation \ref{eqn:est2}
under different values of the  cost budget (the 1\textsuperscript{st} column), while optimizing the completion times.
The 5\textsuperscript{th} column of Table \ref{table:minimize1} gives the completion times $T_{Est}$ estimated using OptEx, and the last
column gives the recorded completion time $T_{Rec}$ with the estimated cluster composition. 

$\newcommand{\Var}{\mathrm{Var}}$
\subsection{Confidence Under Varying Choice of Representative Jobs}\label{sec:conf}
   The job completion times depend on the job profile generated using representative
jobs. 
 With the assumptions regarding the choice of application category (Section \ref{sec:setup}), Table
 \ref{table:confidence} gives the mean, standard deviation, variance, and 95\% confidence intervals for the
 estimated completion time $\mathit{T_{Est}}$, under varying choice of representative jobs. 
 The function $T_{Est}$ is a nonlinear function
over the integer variable $n$, i.e., $\mathit{T_{Est}} = f(n)$ (Section \ref{sec:cost}). 
Let $\mu$ and $\sigma$ be the sample mean and
variance of the job completion times for the experiments, under varying choice of representative job for each category, given in Table
 \ref{table:confidence}. The standard deviation and variance (Table
 \ref{table:confidence}) of the
function represents the stability of the function $f$, under given variations in the choice of representative job. The expectation
and variance (Table
 \ref{table:confidence}) is computed using Taylor expansions \cite{leader2004numerical}, and can be expressed as follows:
E[$f(n)$] $\approx(\mu)$ + $\frac{f''(\mu)}{2}(\sigma)^2$, and Var[$\mathit{T_{Est}}$] $\approx (f'(E[n]))^2$.
The confidence interval acts as a tolerance bound that limits the estimated values of
$\mathit{T_{Est}}$ within an acceptable range.
 We say that as long as 95\% of the estimated $\mathit{T_{Est}}$ values remain
within the interval $\mu - \alpha\sigma$ to $\mu + \alpha\sigma$ (Table
 \ref{table:confidence}), the estimation is acceptable
with 95\% confidence level.

\section{Related Work}\label{sec:related}
    The use of job profiling, performance modeling \cite{
brebner2011}, and benchmarking techniques \cite{Dejun:2011:RPW:2002168.2002173} for efficient load
balancing \cite{daniel2011prediction, Imai:2013:ARP:2588611.2588688}, cost \cite{singer2010towards} and power optimization, have been attempted in quite a few cloud based systems.
There has been considerable amount of work
\cite{Herodotou:2011:NOS:2038916.2038934,Verma:2011:AAR:1998582.1998637, 
alrokayan2014sla
}
 on job scheduling and resource allocation on Hadoop. 
 Verma et. al. \cite{Verma:2011:AAR:1998582.1998637} presents the design of ARIA, a
framework for optimal resource allocation for Hadoop MapReduce.
ARIA \cite{Verma:2011:AAR:1998582.1998637} models Hadoop MapReduce job execution, and cannot be readily applied to other parallel processing frameworks because: 1) it uses Hadoop-specific configuration parameters like map-reduce slots,
 and 2) it
constructs job profile with Hadoop-specific statistics, like running time of map, reduce, and shuffle phases.
 Elastisizer \cite{Herodotou:2011:NOS:2038916.2038934} uses expensive search based or black box based techniques, that require a huge database, for provisioning Hadoop clusters, and has inherent problems like over predicting. 
  Though Spark \cite{Zaharia:2012:RDD:2228298.2228301} is fast surpassing Hadoop in popularity and usage
, there has not been much work in modelling Spark jobs yet.  
\section{Conclusions}\label{sec:conclusion}
OptEx models Spark job execution using analytical techniques. 
    OptEx provides a mean relative error of 6\% in estimating job completion time.
    OptEx yields a success rate of 98\% in completing Spark jobs under a given SLO deadline
    with cost optimal cluster composition estimated using OptEx. OptEx can be used to estimate whether a given job will finish under a given deadline with the given resources on the cloud. It can be used to devise optimal scheduling strategy for Spark. 
\renewcommand{\bibfont}{\footnotesize}
\bibliographystyle{IEEEtran}
\bibliography{cloud}

\begin{thebibliography}{10}
\providecommand{\url}[1]{#1}
\csname url@samestyle\endcsname
\providecommand{\newblock}{\relax}
\providecommand{\bibinfo}[2]{#2}
\providecommand{\BIBentrySTDinterwordspacing}{\spaceskip=0pt\relax}
\providecommand{\BIBentryALTinterwordstretchfactor}{4}
\providecommand{\BIBentryALTinterwordspacing}{\spaceskip=\fontdimen2\font plus
\BIBentryALTinterwordstretchfactor\fontdimen3\font minus
  \fontdimen4\font\relax}
\providecommand{\BIBforeignlanguage}[2]{{%
\expandafter\ifx\csname l@#1\endcsname\relax
\typeout{** WARNING: IEEEtran.bst: No hyphenation pattern has been}%
\typeout{** loaded for the language `#1'. Using the pattern for}%
\typeout{** the default language instead.}%
\else
\language=\csname l@#1\endcsname
\fi
#2}}
\providecommand{\BIBdecl}{\relax}
\BIBdecl

\bibitem{inc_amazon_2008}
\BIBentryALTinterwordspacing
A.~Inc, \emph{Amazon Elastic Compute Cloud {(Amazon} {EC2)}}.\hskip 1em plus
  0.5em minus 0.4em\relax http://aws.amazon.com/ec2/\#pricing: Amazon Inc.,
  2008. [Online]. Available: \url{http://aws.amazon.com/ec2/\#pricing}
\BIBentrySTDinterwordspacing

\bibitem{daniel2011prediction}
\BIBentryALTinterwordspacing
S.~Daniel and M.~Kwon, ``Prediction-based virtual instance migration for
  balanced workload in the cloud datacenters,'' 2011. [Online]. Available:
  \url{http://scholarworks.rit.edu/article/985}
\BIBentrySTDinterwordspacing

\bibitem{Imai:2013:ARP:2588611.2588688}
\BIBentryALTinterwordspacing
S.~Imai, T.~Chestna, and C.~A. Varela, ``Accurate resource prediction for
  hybrid iaas clouds using workload-tailored elastic compute units,'' in
  \emph{Proceedings of the 2013 IEEE/ACM 6th International Conference on
  Utility and Cloud Computing}, ser. UCC '13.\hskip 1em plus 0.5em minus
  0.4em\relax Washington, DC, USA: IEEE Computer Society, 2013, pp. 171--178.
  [Online]. Available: \url{http://dx.doi.org/10.1109/UCC.2013.40}
\BIBentrySTDinterwordspacing

\bibitem{singer2010towards}
G.~Singer, I.~Livenson, M.~Dumas, S.~N. Srirama, and U.~Norbisrath, ``Towards a
  model for cloud computing cost estimation with reserved instances,''
  \emph{Proc. of 2nd Int. ICST Conf. on Cloud Computing, CloudComp 2010}.

\bibitem{Zaharia:2012:RDD:2228298.2228301}
\BIBentryALTinterwordspacing
M.~Zaharia, M.~Chowdhury, T.~Das, A.~Dave, J.~Ma, M.~McCauley, M.~J. Franklin,
  S.~Shenker, and I.~Stoica, ``Resilient distributed datasets: A fault-tolerant
  abstraction for in-memory cluster computing,'' in \emph{Proceedings of the
  9th USENIX Conference on Networked Systems Design and Implementation}, ser.
  NSDI'12.\hskip 1em plus 0.5em minus 0.4em\relax Berkeley, CA, USA: USENIX
  Association, 2012, pp. 2--2. [Online]. Available:
  \url{http://dl.acm.org/citation.cfm?id=2228298.2228301}
\BIBentrySTDinterwordspacing

\bibitem{Herodotou:2011:NOS:2038916.2038934}
\BIBentryALTinterwordspacing
H.~Herodotou, F.~Dong, and S.~Babu, ``No one (cluster) size fits all: Automatic
  cluster sizing for data-intensive analytics,'' in \emph{Proceedings of the
  2Nd ACM Symposium on Cloud Computing}, ser. SOCC '11.\hskip 1em plus 0.5em
  minus 0.4em\relax New York, NY, USA: ACM, 2011, pp. 18:1--18:14. [Online].
  Available: \url{http://doi.acm.org/10.1145/2038916.2038934}
\BIBentrySTDinterwordspacing

\bibitem{Verma:2011:AAR:1998582.1998637}
\BIBentryALTinterwordspacing
A.~Verma, L.~Cherkasova, and R.~H. Campbell, ``Aria: Automatic resource
  inference and allocation for mapreduce environments,'' in \emph{Proceedings
  of the 8th ACM International Conference on Autonomic Computing}, ser. ICAC
  '11.\hskip 1em plus 0.5em minus 0.4em\relax New York, NY, USA: ACM, 2011, pp.
  235--244. [Online]. Available:
  \url{http://doi.acm.org/10.1145/1998582.1998637}
\BIBentrySTDinterwordspacing

\bibitem{Chi:2011:IIC:2002938.2002942}
\BIBentryALTinterwordspacing
Y.~Chi, H.~J. Moon, and H.~Hacig\"{u}m\"{u}\c{s}, ``icbs: Incremental
  cost-based scheduling under piecewise linear slas,'' \emph{Proc. VLDB
  Endow.}, vol.~4, no.~9, pp. 563--574, Jun. 2011. [Online]. Available:
  \url{http://dx.doi.org/10.14778/2002938.2002942}
\BIBentrySTDinterwordspacing

\bibitem{Zhang:2011:TCS:2095686.2095687}
\BIBentryALTinterwordspacing
N.~Zhang, J.~Tatemura, J.~M. Patel, and H.~Hacig\"{u}m\"{u}\c{s}, ``Towards
  cost-effective storage provisioning for dbmss,'' \emph{Proc. VLDB Endow.},
  vol.~5, no.~4, pp. 274--285, Dec. 2011. [Online]. Available:
  \url{http://dx.doi.org/10.14778/2095686.2095687}
\BIBentrySTDinterwordspacing

\bibitem{leader2004numerical}
\BIBentryALTinterwordspacing
J.~Leader, \emph{Numerical Analysis and Scientific Computation}.\hskip 1em plus
  0.5em minus 0.4em\relax Pearson Addison Wesley, 2004. [Online]. Available:
  \url{http://books.google.com/books?id=y-XEGAAACAAJ}
\BIBentrySTDinterwordspacing

\bibitem{apache:library1234}
\BIBentryALTinterwordspacing
A.~S. Foundation, ``Apache spark libraries,'' 2015, [Online; accessed
  25-June-2015]. [Online]. Available: \url{https://spark.apache.org/}
\BIBentrySTDinterwordspacing

\bibitem{brebner2011}
\BIBentryALTinterwordspacing
P.~Brebner and A.~Liu, ``\BIBforeignlanguage{English}{Performance and cost
  assessment of cloud services},'' in
  \emph{\BIBforeignlanguage{English}{Service-Oriented Computing}}, ser. Lecture
  Notes in Computer Science, E.~Maximilien, G.~Rossi, S.-T. Yuan, H.~Ludwig,
  and M.~Fantinato, Eds.\hskip 1em plus 0.5em minus 0.4em\relax Springer Berlin
  Heidelberg, 2011, vol. 6568, pp. 39--50. [Online]. Available:
  \url{http://dx.doi.org/10.1007/978-3-642-19394-1_5}
\BIBentrySTDinterwordspacing

\bibitem{Zafarani+Liu:2009}
\BIBentryALTinterwordspacing
R.~Zafarani and H.~Liu, ``Social computing data repository at {ASU},'' 2009.
  [Online]. Available: \url{http://socialcomputing.asu.edu}
\BIBentrySTDinterwordspacing

\bibitem{MOVIELENS-DATA}
\BIBentryALTinterwordspacing
\emph{{MovieLens dataset, http://www.grouplens.org/data/}}, as of 2003.
  [Online]. Available: \url{http://www.grouplens.org/data/}
\BIBentrySTDinterwordspacing

\bibitem{amplab:benchmark7890}
\BIBentryALTinterwordspacing
AMPLAB, ``Big data benchmark by amplab of uc berkeley,'' 2013, [Online;
  accessed 25-June-2015]. [Online]. Available:
  \url{https://amplab.cs.berkeley.edu/benchmark/}
\BIBentrySTDinterwordspacing

\bibitem{Zuleger:2011:BAI:2041552.2041574}
\BIBentryALTinterwordspacing
F.~Zuleger, S.~Gulwani, M.~Sinn, and H.~Veith, ``Bound analysis of imperative
  programs with the size-change abstraction,'' in \emph{Proceedings of the 18th
  International Conference on Static Analysis}, ser. SAS'11.\hskip 1em plus
  0.5em minus 0.4em\relax Berlin, Heidelberg: Springer-Verlag, 2011, pp.
  280--297. [Online]. Available:
  \url{http://dl.acm.org/citation.cfm?id=2041552.2041574}
\BIBentrySTDinterwordspacing

\bibitem{snapnets}
J.~Leskovec and A.~Krevl, ``{SNAP Datasets}: {Stanford} large network dataset
  collection,'' \url{http://snap.stanford.edu/data}, Jun. 2014.

\bibitem{Pavlo:2009:CAL:1559845.1559865}
\BIBentryALTinterwordspacing
A.~Pavlo, E.~Paulson, A.~Rasin, D.~J. Abadi, D.~J. DeWitt, S.~Madden, and
  M.~Stonebraker, ``A comparison of approaches to large-scale data analysis,''
  in \emph{Proceedings of the 2009 ACM SIGMOD International Conference on
  Management of Data}, ser. SIGMOD '09.\hskip 1em plus 0.5em minus 0.4em\relax
  New York, NY, USA: ACM, 2009, pp. 165--178. [Online]. Available:
  \url{http://doi.acm.org/10.1145/1559845.1559865}
\BIBentrySTDinterwordspacing

\bibitem{YKfJ07}
``{The YourKit Java Profiler},'' http://www.yourkit.com, last 2008.

\bibitem{spark:usecase1234}
\BIBentryALTinterwordspacing
A.~S. Foundation, ``Powered by spark,'' 2015, [Online; accessed
  02-November-2015]. [Online]. Available:
  \url{https://cwiki.apache.org/confluence/display/SPARK/Powered+By+Spark}
\BIBentrySTDinterwordspacing

\bibitem{spark:usecase5678}
\BIBentryALTinterwordspacing
Cloudera, ``Category archives: Use case,'' 2015, [Online; accessed
  02-November-2015]. [Online]. Available:
  \url{http://blog.cloudera.com/blog/category/use-case/}
\BIBentrySTDinterwordspacing

\bibitem{spark:usecase9012}
\BIBentryALTinterwordspacing
S.~Pappas, ``9 super-cool uses for supercomputers,'' 2015, [Online; accessed
  02-November-2015]. [Online]. Available:
  \url{http://www.livescience.com/6392-9-super-cool-supercomputers.html}
\BIBentrySTDinterwordspacing

\bibitem{HadoopYARN}
\BIBentryALTinterwordspacing
A.~S. Foundation. (2014) Apache hadoop nextgen mapreduce (yarn). [Online].
  Available:
  \url{http://hadoop.apache.org/docs/current/hadoop-yarn/hadoop-yarn-site/YARN.html}
\BIBentrySTDinterwordspacing

\bibitem{Dejun:2011:RPW:2002168.2002173}
\BIBentryALTinterwordspacing
J.~Dejun, G.~Pierre, and C.-H. Chi, ``Resource provisioning of web applications
  in heterogeneous clouds,'' in \emph{Proceedings of the 2Nd USENIX Conference
  on Web Application Development}, ser. WebApps'11.\hskip 1em plus 0.5em minus
  0.4em\relax Berkeley, CA, USA: USENIX Association, 2011, pp. 5--5. [Online].
  Available: \url{http://dl.acm.org/citation.cfm?id=2002168.2002173}
\BIBentrySTDinterwordspacing

\bibitem{alrokayan2014sla}
M.~Alrokayan, A.~Vahid~Dastjerdi, and R.~Buyya, ``Sla-aware provisioning and
  scheduling of cloud resources for big data analytics,'' in \emph{Cloud
  Computing in Emerging Markets (CCEM), 2014 IEEE International Conference
  on}.\hskip 1em plus 0.5em minus 0.4em\relax IEEE, 2014, pp. 1--8.

\end{thebibliography}


\begin{thebibliography}{10}

\bibitem{brebner2011}
P.~Brebner and A.~Liu.
\newblock Performance and cost assessment of cloud services.
\newblock In E.~Maximilien, G.~Rossi, S.-T. Yuan, H.~Ludwig, and M.~Fantinato,
  editors, {\em Service-Oriented Computing}, volume 6568 of {\em Lecture Notes
  in Computer Science}, pages 39--50. Springer Berlin Heidelberg, 2011.

\bibitem{4770558}
J.~Choi, S.~Govindan, B.~Urgaonkar, and A.~Sivasubramaniam.
\newblock Profiling, prediction, and capping of power consumption in
  consolidated environments.
\newblock In {\em Modeling, Analysis and Simulation of Computers and
  Telecommunication Systems, 2008. MASCOTS 2008. IEEE International Symposium
  on}, pages 1--10, Sept 2008.

\bibitem{Cooper:2010:BCS:1807128.1807152}
B.~F. Cooper, A.~Silberstein, E.~Tam, R.~Ramakrishnan, and R.~Sears.
\newblock Benchmarking cloud serving systems with ycsb.
\newblock In {\em Proceedings of the 1st ACM Symposium on Cloud Computing},
  SoCC '10, pages 143--154, New York, NY, USA, 2010. ACM.

\bibitem{daniel2011prediction}
S.~Daniel and M.~Kwon.
\newblock Prediction-based virtual instance migration for balanced workload in
  the cloud datacenters.
\newblock 2011.

\bibitem{Dejun:2011:RPW:2002168.2002173}
J.~Dejun, G.~Pierre, and C.-H. Chi.
\newblock Resource provisioning of web applications in heterogeneous clouds.
\newblock In {\em Proceedings of the 2Nd USENIX Conference on Web Application
  Development}, WebApps'11, pages 5--5, Berkeley, CA, USA, 2011. USENIX
  Association.

\bibitem{Gridmix3}
dekel.
\newblock Gridmix3 emulating production workload for apache hadoop.
\newblock April 2010.

\bibitem{Ganesh:2007:OPC:1361397.1361406}
L.~Ganesh, H.~Weatherspoon, M.~Balakrishnan, and K.~Birman.
\newblock Optimizing power consumption in large scale storage systems.
\newblock In {\em Proceedings of the 11th USENIX Workshop on Hot Topics in
  Operating Systems}, HOTOS'07, pages 9:1--9:6, Berkeley, CA, USA, 2007. USENIX
  Association.

\bibitem{hamzahresource}
A.~A. Hamzah, S.~Khattab, and S.~S. El-Gamal.
\newblock Resource allocation for antivirus cloud appliances.

\bibitem{Imai_accurateresource}
S.~Imai, T.~Chestna, and C.~A. Varela.
\newblock Accurate resource prediction for hybrid iaas clouds using
  workload-tailored elastic compute units.

\bibitem{Mann47}
H.~B. Mann and D.~R. Whitney.
\newblock On a test of whether one of two random variables is stochastically
  larger than the other.
\newblock {\em Annals of Mathematical Statistics}, 18(1):50--60, 1947.

\bibitem{Quinlan:1986:IDT:637962.637969}
J.~R. Quinlan.
\newblock Induction of decision trees.
\newblock {\em Mach. Learn.}, 1(1):81--106, Mar. 1986.

\bibitem{Ren12hadoopsadolescence}
K.~Ren, Y.~Kwon, M.~Balazinska, and B.~Howe.
\newblock Hadoop’s adolescence: A comparative workload analysis from three
  research clusters, 2012.

\bibitem{10.1109/TSC.2013.40}
Z.~Ren, J.~Wan, W.~Shi, X.~Xu, and M.~Zhou.
\newblock Workload analysis, implications, and optimization on a production
  hadoop cluster: A case study on taobao.
\newblock {\em IEEE Transactions on Services Computing}, 7(2):307--321, 2014.

\bibitem{Rizvandi:2012:MPT:2403514.2403545}
N.~B. Rizvandi, J.~Taheri, R.~Moraveji, and A.~Y. Zomaya.
\newblock On modelling and prediction of total cpu usage for applications in
  mapreduce environments.
\newblock In {\em Proceedings of the 12th International Conference on
  Algorithms and Architectures for Parallel Processing - Volume Part I},
  ICA3PP'12, pages 414--427, Berlin, Heidelberg, 2012. Springer-Verlag.

\bibitem{singer2010towards}
G.~Singer, I.~Livenson, M.~Dumas, S.~N. Srirama, and U.~Norbisrath.
\newblock Towards a model for cloud computing cost estimation with reserved
  instances.
\newblock {\em Proc. of 2nd Int. ICST Conf. on Cloud Computing, CloudComp
  2010}, 2010.

\bibitem{uddincloud}
M.~Uddin, B.~He, and R.~Sion.
\newblock Cloud performance benchmark series, amazon relational database
  service (rds) tpc-c benchmark.
\newblock Technical report, Technical report, Stony Brook Network Security and
  Applied Cryptography Lab.

\bibitem{Wang:2011:TSR:2060110.2060857}
G.~Wang, A.~R. Butt, H.~Monti, and K.~Gupta.
\newblock Towards synthesizing realistic workload traces for studying the
  hadoop ecosystem.
\newblock In {\em Proceedings of the 2011 IEEE 19th Annual International
  Symposium on Modelling, Analysis, and Simulation of Computer and
  Telecommunication Systems}, MASCOTS '11, pages 400--408, Washington, DC, USA,
  2011. IEEE Computer Society.

\bibitem{wiki:roc}
Wikipedia.
\newblock Receiver operating char. --- wikipedia{,} the free encyclopedia,
  2014.
\newblock [Online; accessed 17-July-2014].

\bibitem{Yanggratoke:2011:GRA:2147671.2147698}
R.~Yanggratoke, F.~Wuhib, and R.~Stadler.
\newblock Gossip-based resource allocation for green computing in large clouds.
\newblock In {\em Proceedings of the 7th International Conference on Network
  and Services Management}, CNSM '11, pages 171--179, Laxenburg, Austria,
  Austria, 2011. International Federation for Information Processing.

\end{thebibliography}

\end{document}